\documentclass[pra,aps,superscriptaddress,footinbib,twocolumn]{revtex4-1}

\usepackage{mathrsfs}
\usepackage{amsfonts}
\usepackage{amssymb}
\usepackage{multirow}
\usepackage{svg}
\usepackage{float}
\makeatletter
\let\newfloat\newfloat@ltx
\makeatother
\usepackage{algorithm}
\usepackage{algpseudocode}
\usepackage{graphicx}
\usepackage{subfigure}
\usepackage{amsmath} 

\usepackage[colorlinks=true,citecolor=blue,linkcolor=magenta]{hyperref}

\usepackage{lmodern}

\usepackage{dsfont}

\usepackage{natbib}

\usepackage{bm}

\usepackage{pgf}

\usepackage{calligra}
\usepackage{qcircuit}

\usepackage{threeparttable}

\usepackage{amsmath}

\usepackage{tabu}

\usepackage{appendix}

\newcommand{\Tr}{\mathrm{Tr}}

\newcommand{\ket}[1]{\vert{ #1 }\rangle}

\newcommand{\bra}[1]{\langle{ #1 }\vert}

\begin{document}

\title{Scalable Algorithms for calculating
Power Function of Random Quantum States in NISQ Era}
\author{Wencheng Zhao}
\altaffiliation{These authors contributed equally to the work.}
\affiliation{China University of Mining and Technology, College of Sciences, Beijing 100083, China}

\author{Tingting Chen}
\altaffiliation{These authors contributed equally to the work.}
\affiliation{China University of Mining and Technology, College of Sciences, Beijing 100083, China}

\author{Ruyu Yang}
\email{yangruyu96@gmail.com}
\affiliation{Graduate School of China Academy of Engineering Physics, Beijing 100193, China}

\begin{abstract}
This article focuses on the development of scalable and quantum bit-efficient algorithms for computing power functions of random quantum states. Two algorithms, based on Hadamard testing and Gate Set Tomography, are proposed. We provide a comparative analysis of their computational outcomes, accompanied by a meticulous evaluation of inherent errors in the gate set tomography approach. The second algorithm exhibits a significant reduction in the utilization of two-qubit gates compared to the first. As an illustration, we apply both methods to compute the Von Neumann entropy of randomly generated quantum states.
\end{abstract}
\maketitle

\section{Introduction}\label{introduction}


 Random quantum states form the foundational basis for our understanding of Quantum Information\cite{swingle2016measuring,brandao2021models}, Black holes\cite{hayden2007black,kudler2021relative}, and related fields.  Numerous important functions, such as Renyi entropy, Von Neumann entropy, Quantum Fisher information, fidelity of random states, virtual distillation, and separation of density matrices\cite{holmes2023nonlinear,subramanian2021quantum,zyczkowski2006introduction,rath2021quantum,jozsa1994fidelity,koczor2021exponential,wang2022new}, play crucial roles in quantum information, condensed matter physics, quantum chemistry, and beyond\cite{kandala2017hardware,carteret2005noiseless,lubasch2020variational,georgeot2001exponential,elben2020many,braumuller2022probing}. Quantum computing holds a significant computational efficiency advantage over classical computing\cite{national2019quantum}. The current development of quantum devices is situated in the Noisy Intermediate-Scale Quantum (NISQ) era, characterized by the handling of qubits in the tens or hundreds, accompanied by inevitable quantum noise\cite{preskill2018quantum,lau2022nisq,ding2022quantum}. Exploiting the advantages and addressing the challenges of NISQ quantum computers, we tackle the fundamental yet challenging task of developing algorithms for computing nonlinear functions of random quantum state.
 

Prior methodologies for nonlinear transformations relied on simultaneously preparing multiple copies of a quantum state\cite{zhou2022hybrid,holmes2023nonlinear,bovino2005direct,horodecki2003measuring,ekert2002direct} and collective measurements\cite{bovino2005direct,ekert2002direct,horodecki2003measuring}. These approaches necessitated a large number of qubits. For instance, when computing $\Tr\{\rho^m\}$, with $\rho$ representing the density matrix defined over $n$ qubits, these methods required $nm$ qubits. However, in the NISQ era, the number of qubits is still insufficient, rendering it inadequate to achieve quantum advantage within these algorithms\cite{preskill2018quantum}.
Conversely, researchers have advocated for constructing the classical shadow of $\rho$ and subsequently employing it to compute the purity $\Tr\{\rho^2\}$\cite{sack2022avoiding,zhang2021experimental,seif2023shadow,elben2023randomized,brydges2019probing,elben2019statistical,elben2020mixed}. While this approach still entails exponential resources relative to the number of qubits, it is perceived as an improvement over traditional State Tomography\cite{o2016efficient}. Nevertheless, ongoing exploration of such methods is delimited to purity, which corresponds to quadratic functions of the density matrix. For higher-order functions like $\Tr\{\rho^m\}$, there is no verified indication that these methods sustain an advantage over classical approaches.

To more efficiently exploit quantum computers in the NISQ era, we aim to design algorithms that employ the same number of qubits as $\rho$, and the circuit depth exhibits polynomial growth with the order of nonlinear function. A technique for generating random states involves initiating from an initial state and applying quantum gates randomly based on a specific probability distribution. The resultant final states post the application of diverse quantum gates to the initial state might not be orthogonal. We ascertain the presence of the algorithm we want, assuming the knowledge about how to construct the intended random state by utilizing random circuits.

In this study, we introduce two distinct algorithms, both characterized by their shared utilization of the Grover gate $G=I-2|0\rangle\langle0|$. The primary aim of both algorithms is to compute the power series expansion $\Tr\{\rho^m\}$ for a nonlinear function in the context of a multi-qubit quantum random state $\rho$. The first algorithm is based on the Hadamard Test(HT). It involves transforming an auxiliary qubit (usually $|0\rangle\langle0|$) into a superposition state using the Hadamard gate. After a controlled gate operation, another Hadamard gate extracts essential data, finalizing the calculation. Our algorithm is Hadamard Test-based but introduces an innovative approach:  we deploy a quantum pure state circuit to simulate $\Tr\{G^{m-1}\rho\}$ computation for a quantum random state, by employing weighted averages across multiple measurements.
The second algorithm begins by mathematically converting the calculation of $\Tr\{\rho^m\}$ for the desired quantum state into $\Tr\{G^m\}$. A comprehensive understanding of $G^m$ is acquired through Gate Set Tomography(GST)\cite{d2003quantum,yang2021perturbative} in a subspace, facilitating the calculation and estimation of $\Tr\{G^m\}$ through mathematical processing. Compared to the Hadamard Test-based algorithm, this approach entails fewer qubits and two-qubit gates within the circuit. Moreover, this method introduces a novel result-processing technique rooted in reconstruction. Both of these algorithms are scalable, with their time complexity growing polynomially with the number of qubits.


The structure of this paper is outlined as follows. We begin by introducing the Hadamard test-based algorithm in Section \ref{Hadamard-test}. Subsequently, we explain the application of the GST method in Section \ref{Tomography} for extracting relevant information from the subspace. Section \ref{Error Analysis} is devoted to error analysis and complexities. In Section \ref{results-and-discussion}, we present computational results garnered from the preparation of the random quantum state $\rho$ and the subsequent application of both algorithms. Notably, we compare the variations in calculations for the identical quantum state when utilizing the two distinct algorithms. A concise summary of this article is offered in Section \ref{conclusion}.

In the appendix, we explore into specific scenarios, enabling the applicability of our algorithms to various instances of solving nonlinear functions within random quantum states. Additionally, we conduct result modeling under the presence of noise to clarify the impact of noise-induced uncertainty on outcomes. Furthermore, we also investigate the dimensions of the subspaces that need to be studied when applying our algorithms to more general functions.
\section{Trace Estimation of Hadamard Test}\label{Hadamard-test}
\subsection{Theoretical Part of Hadamard Test}
In this section, we show how to calculate $\Tr\{\rho^m\}$ using HT\cite{wu2021quantum,xu2022variational,patti2023quantum,aharonov2006polynomial}. Firstly, we show how the quantum state is encoded in a quantum channel.

Suppose the $n$-qubits random quantum state is given by $\mathrm{\rho}=\sum_{i=1}^\alpha {p}_iU_i|0\rangle\langle0|^{\otimes n} U_i^\dagger=\sum_{i=1}^\alpha p_i|\psi_i\rangle\langle\psi_i|$,
where $\alpha$ random unitary gates $U_i$ and probabilities $p_i$ are known.
Define the $G$ gate as: $G=\sum_{j=1}^{\alpha}p_j U_jG_0 U_j^\dagger$, where $G_0=I_{2^n}-2|0\rangle\langle0|^{\otimes n }$ is the Grover operator. Then we can express the quantum channel as $G=I_{2^n}-2\rho$, and $G^k=(I_{2^n}-2\rho)^k$. In this way, We encode the state $\rho$ into a non-unitary quantum channel $G$.

\begin{figure}[b]
        \centering
        \includegraphics[width=1.0\linewidth]{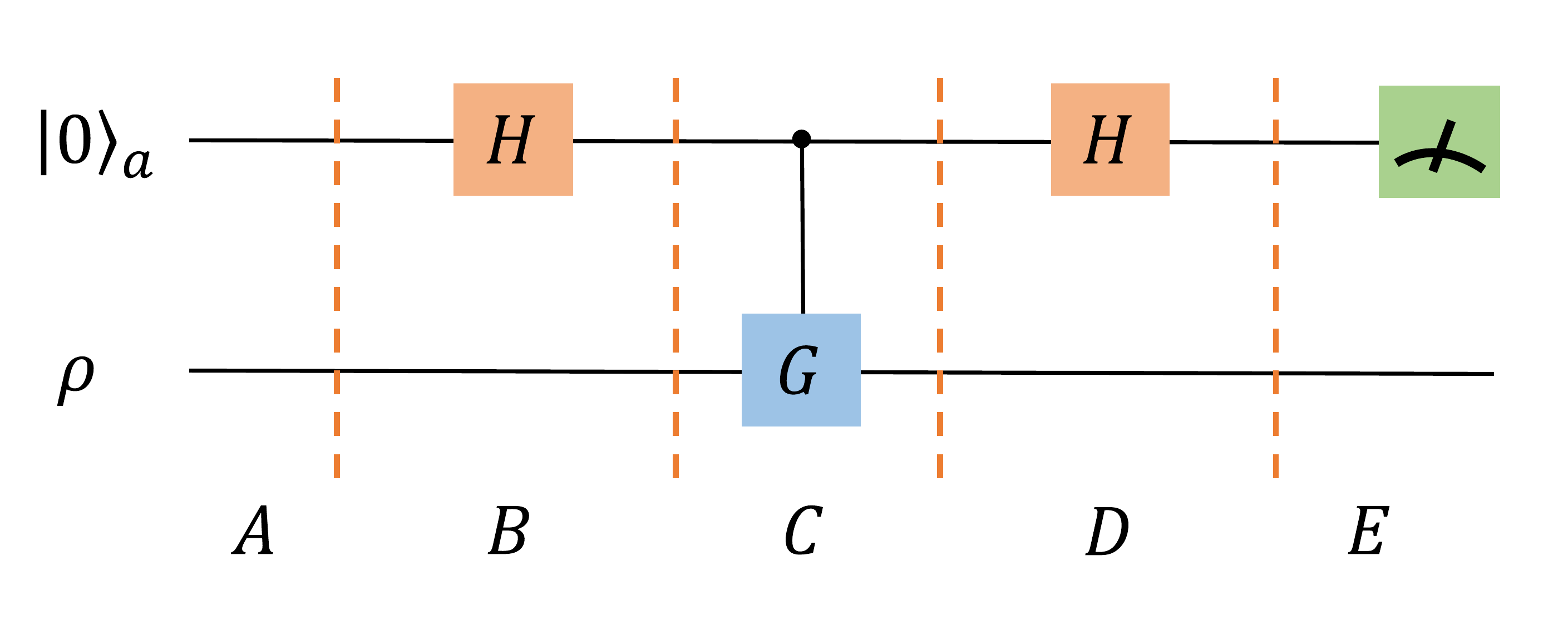}
        \caption{Illustrative diagram of the Hadamard Test circuit. The diagram features two distinct paths, which have been divided into five segments for ease of computation. The first path encompasses two $H$ gates and a single measurement gate, while the second path incorporates a $controlled-G$ gate.}
    \label{fig:1}
\end{figure}

Next, we show how HT works. In general, one ancilla qubit is required for HT.
The quantum circuit has been shown in Fig~\ref{fig:1}. The computation process is as follows. 

In the region denoted as \(\mathrm{A}\), the quantum state of the circuit at this point is
given by:
\begin{equation}
|0\rangle\langle0|\otimes\rho.    
\end{equation} 
In the \(\mathrm{B}\) region, after applying the Hadamard gate, the
state of the ancillary qubit is transformed as Eq.(\ref{eq:1}):
\begin{equation}
\mathrm{H}|0\rangle\langle0|\mathrm{H}^\dagger=\frac{1}{2}(|0\rangle\langle0|+|0\rangle\langle1|+|1\rangle\langle0|+|1\rangle\langle1|).
\label{eq:1}
\end{equation}
The overall state of the circuit is given by:
\begin{equation}
\frac{1}{2}(|0\rangle\langle0| + |0\rangle\langle1| + |1\rangle\langle0| + |1\rangle\langle1|) \otimes \rho.
\end{equation}
Then, the state undergoes the action of the controlled $\mathrm{G}$ gate $CG^k$ :
\begin{equation}
CG^k=|0\rangle\langle0|\otimes I+|1\rangle\langle1|\otimes G^k.
\end{equation}
Thus, the state in the region $\mathrm{C}$ is:
\begin{equation}
\begin{split}
&\frac{1}{2}(|0\rangle\langle0|\otimes\rho)+\frac{1}{2}|0\rangle\langle1|\otimes \rho (G^k)^\dagger)+\\
&\frac{1}{2}|1\rangle\langle0|\otimes G^k\rho)+\frac{1}{2}|1\rangle\langle1|G^k\rho (G^k)^\dagger.
\end{split}
\end{equation}

Subsequently, the auxiliary qubit undergoes another
\(\mathrm{Hadamard}\) gate operation. Consequently, the state  within the region \(\mathrm{D}\) is then given by:
\begin{equation}
\begin{split}
&\frac{1}{2}\left[(H|0\rangle\langle0|H^\dagger)\otimes\rho + (H|0\rangle\langle1|H^\dagger)\otimes \rho (G^k)^\dagger \right. \\
&\left. + (H|1\rangle\langle0|H^\dagger)\otimes G^k\rho + (H|1\rangle\langle1|H^\dagger)\otimes G^k\rho (G^k)^\dagger\right].
\end{split}
\end{equation}
Expanding each term results as:
\begin{equation}
\begin{aligned}
\frac{1}{4}|0\rangle\langle0|\otimes[\rho+\rho (G^k)^\dagger+(G^k)\rho+(G^k)\rho (G^k)^\dagger],\\
\frac{1}{4}|0\rangle\langle1|\otimes[\rho-\rho (G^k)^\dagger+(G^k)\rho-(G^k)\rho (G^k)^\dagger],\\
\frac{1}{4}|1\rangle\langle0|\otimes[\rho+\rho (G^k)^\dagger-(G^k)\rho-(G^k)\rho (G^k)^\dagger],\\
\frac{1}{4}|1\rangle\langle1|\otimes[\rho-\rho (G^k)^\dagger-(G^k)\rho+(G^k)\rho (G^k)^\dagger].
\end{aligned}
\end{equation}
The expression above can be denoted as follows:
\begin{equation}
\rho' = |0\rangle\langle0|\otimes a_{11} + |0\rangle\langle1|\otimes a_{12}+ |1\rangle\langle0|\otimes a_{21}+ |1\rangle\langle1|\otimes a_{22}  
\label{eq:2}
\end{equation}

Under the computational basis measurement in region \(\mathrm{E}\), the
measurement operators are defined as
\(M_0 = |0\rangle\langle0|\otimes I\) and
\(M_1 = |1\rangle\langle1|\otimes I\). 
The probability distribution over the outcomes of the measurement are :
\begin{equation}
\begin{aligned}
{P}(0)&=\Tr\{M_0\rho^\prime M_0^\dagger\}=|a_{11}|^2\\
&=\frac{1}{2}+\frac{1}{2}\Tr\{G^k\rho\},
\label{eq12}
\end{aligned}
\end{equation}

\begin{equation}
\begin{aligned}
{P}(1)&=\Tr\{M_1\rho^\prime M_1^\dagger\}=|a_{22}|^2\\
&=\frac{1}{2}-\frac{1}{2}\Tr\{G^k\rho\}.
\label{eq13}
\end{aligned}
\end{equation}
The expression for $\Tr\{G^k\rho\}$ can be derived, with the ultimate goal of estimating: 
\begin{equation}
\begin{aligned}
    \Tr\{\rho^{m+1}\}& =\Tr\{(\frac{1}{2}I - \frac{1}{2}G)^m\rho\}\\
    &= \frac{1}{2^m}\sum_{k=0}^{m}C_{m}^{k}(-1)^{k}\Tr\{G^k\rho\}.
    \label{eq:15}
\end{aligned}
\end{equation}
Let: $p_k=\frac{1}{2^m}C_{m}^{k}$, $x_k=(-1)^{k}\Tr\{G^k\rho\}$.
The above equation can be transformed into an expectation calculation:

\begin{equation}
\begin{aligned}
    \Tr\{\rho^{m+1}\}=\sum_{k=0}^{m}p_kx_k=E_k(x_k).
\end{aligned}
\end{equation}
To estimate the expectation, we need to generate quantum circuits using a random sampling method. For each circuit, we sample $m$ times, with a $1/2$  probability of adding a $CG$ gate to the circuit and a $1/2$  probability of doing nothing. After generating multiple circuits, we take the average of the results.
\subsection{Algorithm  for calculating $\Tr\{\rho^{m+1}\}$.}

The pseudocode below, referred to as Algorithm $\ref{alg:1}$, is for calculating $\Tr\{\rho^{m+1}\}$.

\begin{algorithm}
\renewcommand{\algorithmicrequire}{\textbf{Input:}}

\renewcommand{\algorithmicensure}{\textbf{Output:$\Tr\{\rho^{m+1}\}$}}
\algorithmicrequire{$n$,$p_i$,$U_i$,$N$}\\

\algorithmicensure{}
\caption{ }
	\label{alg:1}
	\begin{algorithmic}[1]
        \State Set the initial state to $|0\rangle^{\otimes n+1}$.
        \State Randomly select $U_i$ with probability $p_i$ to act on the target qubits.
        \State Apply a Hadamard gate to the ancilla qubit.
        \State Sample $m$ times, with a $1/2$ probability of adding a $CG$ gate to the circuit and a $1/2$ probability of doing nothing.
        \State Apply another Hadamard gate to the ancilla qubit.
        \State Perform a computational basis measurement on the auxiliary qubit circuit. 
        \State Repeat the above steps $N$ times and take the average.
	\end{algorithmic}  
\end{algorithm}

\begin{figure}[b]
        \centering
        \includegraphics[width=1.0\linewidth]{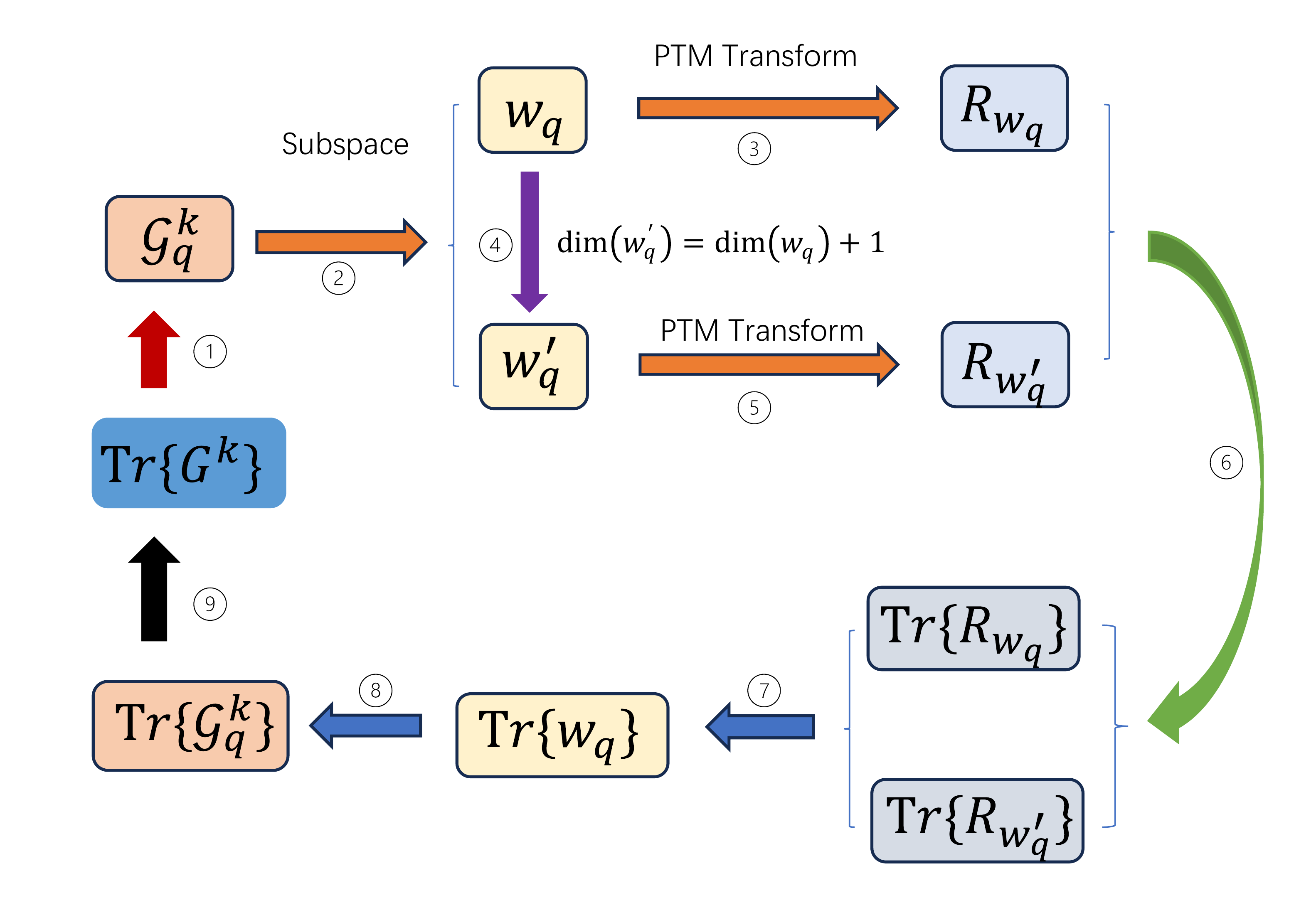}
        \caption{Process of calculating $\Tr\{G^k\}$ using Gate Set Tomography. Firstly, in Step 1, decompose $\Tr\{G^k\}$ into the calculation of each $\Tr\{\mathcal{G}_q^k\}$. Then, take its subspace $w_q$ (Step 2) and transform the matrix representation of the subspace into the corresponding Pauli Transfer Matrix, denoted as $R_{w_q}$. Simultaneously, in Step 4, expand the subspace $w_q$ by one dimension. Next, similarly, transform its matrix representation $w^\prime_q$ into its corresponding Pauli Transfer Matrix, denoted as $R_{w^\prime _q}$. Then, through Step 6, calculate their traces separately. By employing certain mathematical techniques in Step 7, we can use the obtained traces to calculate $\Tr\{w_q\}$, and subsequently obtain $\Tr\{\mathcal{G}_q^k\}$ (Step 8). Finally, in Step 9, we obtain $\Tr\{G^k\}$, which can be seen as the inverse process of Step 1.}
        \label{fig:liucheng}
\end{figure}
\begin{figure*}
        \centering
        \includegraphics[width=1.0\linewidth]{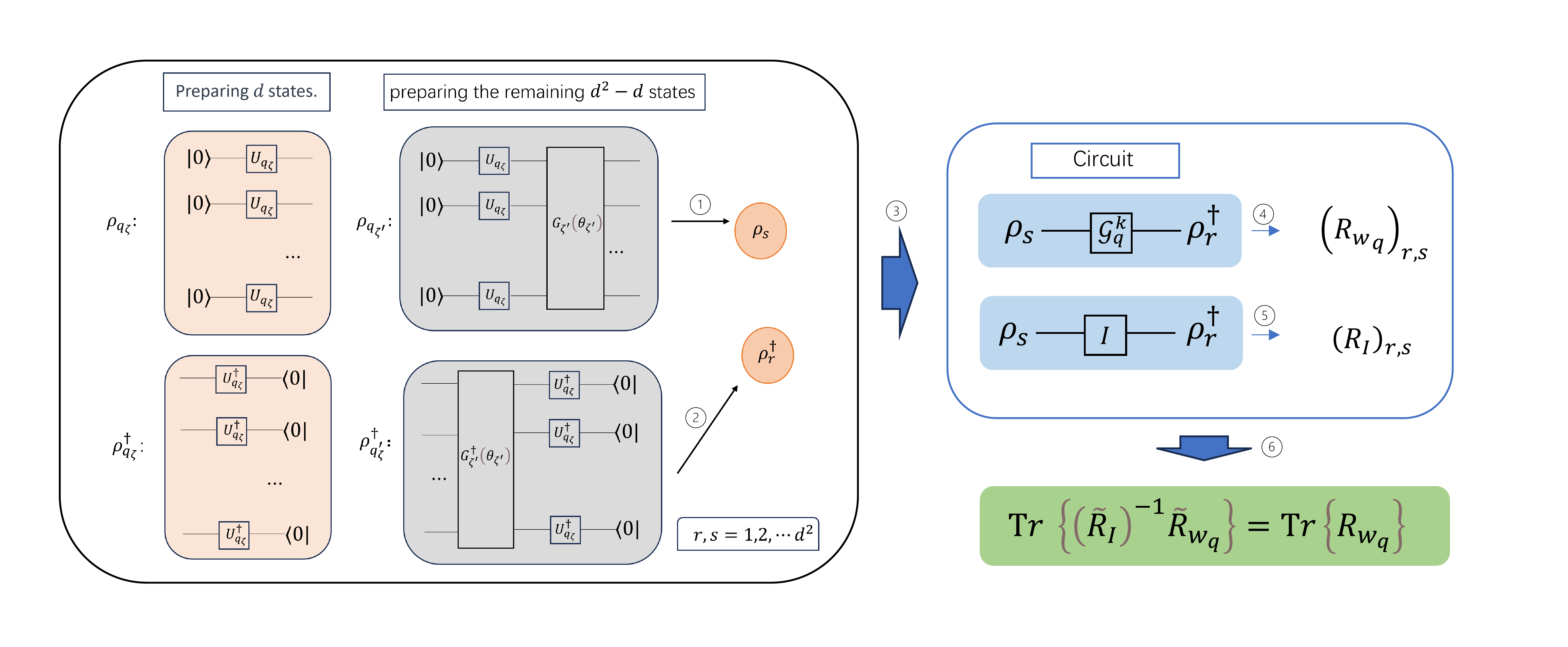}
        \caption{Schematic diagram of a sub-process for computing $\Tr\{R_{w_q}\}$ within the Tomography workflow. The diagram runs from left to right, with the first black box on the left representing the preparation of the initial state $(1)$ and the measurement state $(2)$. The blue box on the upper right depicts the placement of the prepared states into the circuit $(3)$, where the $\mathcal{G}_q^k$ gate and the identity gate are inserted. Steps $(4)$ and $(5)$ allow us to obtain the matrix elements of the PTMs corresponding to each of these gates. Finally, in step $(6)$, $\Tr\{R_{w_q}\}$ is indirectly obtained through a similarity transformation.}
        \label{fig:liucheng1}
\end{figure*}

\section{Trace Estimation of Quantum Tomography}\label{Tomography}





\subsection{Theoretical Part of Quantum Tomography}
Quantum tomography denotes a suite of techniques aiming to reconstruct an unknown quantum channel or state through experimental measurements. This process is pivotal for the comprehensive understanding and authentication of quantum apparatus \cite{greenbaum2015introduction,torlai2018neural,cramer2010efficient,mohseni2008quantum,altepeter2003ancilla}. Nevertheless, the scalability of quantum tomography poses a challenge, as the indispensable measurements and computational resources experience exponential growth in tandem with qubit numbers. Note that although the resources required for tomography can be reduced by representing quantum states as MPS, in the worst case the resources required for tomography still increase exponentially with the number of bits \cite{cramer2010efficient,lanyon2017efficient}. In the context of our current research problem, there is a silver lining: the subspace we are investigating maintains a dimension that remains unaffected by the number of qubits. This distinctive feature becomes particularly advantageous. In the subsequent section, we expound on our utilization of the GST method to extract the pertinent information from this designated subspace. 


In comparison with the preceding context, the process of preparing random quantum states adheres to the same approach as the Hadamard Test method. This methodology necessitates the application of an assortment of stochastically selected gates \(\{(U_i)_{2^n\times 2^n}, i=1,2,\ldots,\alpha\}\) onto the initial quantum state (typically \(|0\rangle\langle0|^{\otimes n}\)), resulting in the emergence of a random quantum state. The primary objective revolves around the computation of \(\Tr\{\rho^m\}\), which is achieved through the intermediary of \(\Tr\{G^m\}\), where \(G=I-2\rho\). We can express $\Tr\{\rho^m\}$ as
\begin{equation}
\Tr\{\rho^m\}=\frac{1}{2^m}\sum_{k=0}^{m}C_{m}^{k}(-1)^{k}\Tr\{G^k\}.   
\end{equation}


Similar to the approach employed in the HT-based algorithm, we can estimate this summation by leveraging the Monte Carlo method. This involves conducting multiple circuit samplings in accordance with their respective probabilities and subsequently calculating the average. Through this process, we can attain the sought-after value of \(\Tr\{\rho^m\}\). The figure shown in Fig~\ref{fig:liucheng} presents our computation process through a simplified flowchart, starting from the decomposition of $\Tr\{G^k\}$, transitioning to the calculation of each individual $\mathcal{G}_q^k$, and then proceeding with a series of computations within their respective subspaces. Finally, the computed results $\Tr\{\mathcal{G}_q^k\}$ are weighted and summed to obtain the desired calculation $\Tr\{G^k\}$. We illustrate the sub-process of computing $\Tr\{R_{w_q}\}$ in Fig~\ref{fig:liucheng1}, which, together with Fig~\ref{fig:liucheng}, forms the complete Tomography computation process.


\subsubsection{Mathematical Treatment:}\label{mathematical-treatment}
We start with the unitary quantum gate \(G_0\):
\begin{equation}
(G_0)_{2^n\times2^n}=I_{2^n}-2|0\rangle\langle0|^{\otimes n}=\begin{pmatrix} -1 & 0 & \cdots & 0 \\
    0 & 1& \cdots & 0 \\
    \vdots & \vdots & \ddots & \vdots \\
    0 & 0 & \cdots & 1\end{pmatrix}.  
\end{equation}
Through the application of random unitary gates \(U_i\) to the initial gate \(G_0\), a set of \(\alpha\) distinct gates is generated. These gates are denoted as \(G_i = U_i G_0 U_i^\dagger\). The composite gate \(G\) is then defined as the weighted summation of these transformed gates: \(G = \sum_{i=1}^{\alpha} p_i G_i\).





An insightful observation can be made that in the presence of \(k\) occurrences of \(G\) gates within the circuit, the total count of possible arrangements aggregates to \(\alpha^k\). These arrangements are uniquely labeled by the index \(q = \{1, 2, \ldots, \alpha^k\}\). As a result, this algorithm effectively dissects the trace \(\Tr\{G^k\}\) of the higher-order powers of \(G\) into computations encompassing \(\alpha^k\) arrangements denoted as \(\mathcal{G}^k_q\). The calculation process is thereby  executed on higher-order random quantum states via the application of a Monte Carlo methodology.


The corresponding
probability combination \(\prod_{t=1}^k p_{q_t}\) is represented as
\(\mathcal{P}_q\).
Therefore, \(\Tr\{G^k\}\) can be expressed as:
\begin{equation}
\Tr\{{G}^k\}=\sum_{q=1}^{\alpha^k}\mathcal{P}_{q}\Tr\{\mathcal{G}_q^k\},
\end{equation}
Where we use
\(\mathcal{G}^k_{q}\) to denote \(\prod_{t=1}^{k}G_{q_t}\)
for convenience.

\subsubsection{The matrix representation of $\mathcal{G}_q^k$}

\label{the-matrix-representation-of-uxagym}


For each of the $\alpha^k$ instances of $\mathcal{G}_q^k$, a specific $\mathcal{G}_q^k$ is chosen for computation where $q=1,2,\ldots,\alpha^k$. In this scenario, the calculation method is provided for arbitrary combinations, while the computation process remains similar for other combinations.


Upon choosing a specific combination \(\mathcal{G}_q^k\), a set of \(k\) corresponding \(G_{q_t}\) gates is determined, thereby giving rise to \(k\) specific \(|\psi_{q_t}\rangle\) states, where \(t=1,2,\ldots,k\).

For example, consider the
cases:
\begin{align*}
&q=1, \{\mathcal{G}_1^k:\underbrace{G_1G_1\cdots G_1G_1}_{k\ layers}\}.\\
&q=2, \{\mathcal{G}_2^k:\underbrace{G_1G_1\cdots G_1G_2}_{k\ layers}\}.\\
& \quad \vdots\\
&q=\alpha^k, \{\mathcal{G}_{\alpha^k}^k:\underbrace{G_\alpha G_\alpha \cdots G_\alpha G_\alpha}_{k\ layers}\}.
\end{align*}
Therefore, under the matrix background denoted as $\mathcal{G}_q^k$, the subspace dimension $d$ is not constant. Hence, the determination of the subspace dimension relies entirely on the count of unique gate types present in the given order $q$. Let us define $d$ as the dimension of the nontrivial subspace corresponding to the simplified merge of $\mathcal{G}_q^k := \prod_{t=1}^{k}G_{q_t}$, while representing the random gate sets used to prepare $G_{q_t}$ as $U_{q_t}$: $G_{q_t}=U_{q_t}G_0U_{q_t}^\dagger$. Quantum states prepared by different random gates are represented as: ${|\psi_{q_\zeta}\rangle, \zeta=1,2,\ldots,d}$.

In order to ensure completeness, a set of state vectors
$\{|\phi_{\eta}\rangle,\eta=1,2,\cdots,2^n-d\}$ is introduced,
which are orthogonal to all the state vectors $|\psi_{q_\zeta}\rangle$.

The rationale behind this is as follows:

Due to the condition \(\langle\psi_{q_\zeta}|\phi_\eta\rangle=0\), it can be
inferred that:
\begin{equation}
\mathcal{G}_q^k(|\phi_\eta\rangle) = \left[\prod_{\zeta=1}^{d}(I-2(|\psi_{q_\zeta}\rangle\langle\psi_{q_\zeta}|)\right]|\phi_\eta\rangle = |\phi_\eta\rangle.  \nonumber
\end{equation}
It is evident that $\{{|\phi_\eta\rangle}, \eta=1,2,\dots,2^n-d\}$ forms a set of eigenstates of $\mathcal{G}_q^k$ with eigenvalue $1$.
In the representation with \(2^n\) state vectors:
\begin{align}
\left\{\underbrace{|\psi_{q_1}\rangle,|\psi_{q_2}\rangle,\ldots,|\psi_{q_d}\rangle}_{d \ \text{terms}},\underbrace{|\phi_1\rangle,|\phi_2\rangle,\ldots,|\phi_{\eta}\rangle }_{(2^n-d) \ \text{terms}}\right\}.
\end{align}
as the basis in the \(V_{q}\) space, the matrix \(\mathcal{G}_q^k\) can be expressed as:
\begin{equation}
\left(
\begin{array}{ccccccc}
w_{11} & w_{12} & \ldots & w_{1d} & 0 & 0 & \ldots \\
w_{21} & w_{22} & \ldots & w_{2d} & 0 & 0 & \ldots \\
\ldots & \ldots & \ldots & \ldots & 0 & 0 & \ldots \\
w_{d1} & w_{d2} & \ldots & w_{dd} & 0 & 0 & \ldots \\
0 & 0 & 0 & 0 & 1 & 0 & \ldots \\
0 & 0 & 0 & 0 & 0 & \ldots & \ldots \\
\ldots & \ldots& \ldots & \ldots & \ldots & \ldots& 1  \\

\end{array}
\right)
\end{equation}
The top-left $d\times d$ matrix $w_q$ can be considered as composed
of the eigenvalues of the $d$-dimensional invariant subspace $V_{q1}$ spanned
by $d$ states $|\psi_{q_\zeta}\rangle$ in the basis. The remaining part of $\mathcal{G}_q^k$ is composed of the
eigenvectors in the complement space with basis ${|\phi_\eta\rangle}$,
corresponding to \((2^n-d)\) eigenvalues. These eigenvalues form an
\((2^n-d)\times (2^n-d)\) identity matrix \(I_{(2^n-d)\times (2^n-d)}\).
Therefore, \(w_q\) is actually the matrix representation of \(\mathcal{G}_q^k\) in the
\(d\)-dimensional invariant subspace $V_{q1}$. Through
this expression, the calculation of \(\Tr\{\mathcal{G}_q^k\}\) can be
performed:
\begin{equation}
\begin{aligned}
\Tr\{\mathcal{G}_q^k\}&=\Tr\{w_{q})+\Tr\{I_{2^n-d}\}\\
&=\Tr\{w_{q}\}+2^n-d.
\end{aligned}
\end{equation}

\subsubsection{\texorpdfstring{The PTM representation of
\(w_q\)}{The PTM representation of w\_q}}

\label{the-ptm-representation-of-uxawy}

After characterizing \(\mathcal{G}_q^k\), we need to obtain the solution for \(\Tr\{w_q\}\). However, the matrix \(w_q\) is unknown. \(w_q\) as a mapping, finding the solution for \(w_q\) requires the use of the Pauli Transfer Matrix(PTM). We denote the PTM corresponding to \(w_q\) as \(R_{w_q}\).

Based on the previous discussion, an important relationship can be used:
\begin{equation}
\begin{aligned}
R_{w_q}|\rho_{s}\rangle\rangle&=|w_q(\rho_{s})\rangle\rangle=|w_q\rho_{s}w_q^\dagger\rangle\rangle\\
&r,s\in\{1,2,\ldots,d^2\}.\nonumber
\end{aligned}
\end{equation}
For simplicity, the subscript \(q\) indicating that \(\rho\) belongs to the ordering
\(\mathcal{G}_q^k\) will be omitted below.

Although \(R_{w_q}\) is an $(d^2\times d^2)$ matrix, the vector
\(|\rho_s\rangle\rangle\) has dimensions of \((2^n)^2\times 1\). Since
\(|\rho_s\rangle\rangle\) only has non-zero elements in the
\(d^2\)-dimensional subspace, the remaining part of the vector is
trivial, consisting of all zeros except for this subspace.

By left-multiplying the above equation by \(\langle\langle \rho_r|\),
the matrix elements of the \(PTM\) matrix \((R_{w_q})_{d^2\times d^2}\)
are given as follows:
\begin{equation}
\langle\langle \rho_r|R_{w_q}|\rho_s\rangle\rangle =\langle\langle\rho_r|w_q\rho_sw_q^\dagger\rangle\rangle.  
\end{equation}
To calculate the matrix elements \((R_{w_q})_{rs}\) of \(R_{w_q}\) using this method, \(d^2\) quantum states  in the Hilbert-Schmidt space are required, represented by vectors
\(|\rho_r \rangle\rangle\).  The proof of the completeness of these states
can be found in Appendix \ref{Appendix A: Completeness Proof}.

We need to calculate the trace of PTM. Although we can obtain each matrix element of the PTM sequentially through the circuit, we cannot directly compute its trace because the basis vectors in the subspace we generate are not orthogonal. There are two feasible approaches:Schmidt decomposition to orthogonalize its basis vectors, and then directly calculate the trace by summing the main diagonal elements;Calculate its trace indirectly through a matrix similarity transformation.This paper adopts the GST (Gate Set Tomography) method, which is the latter approach of indirectly calculating the trace of PTM through similarity transformations.

\begin{figure}[b]
        \centering
        \includegraphics[width=1.0\linewidth]{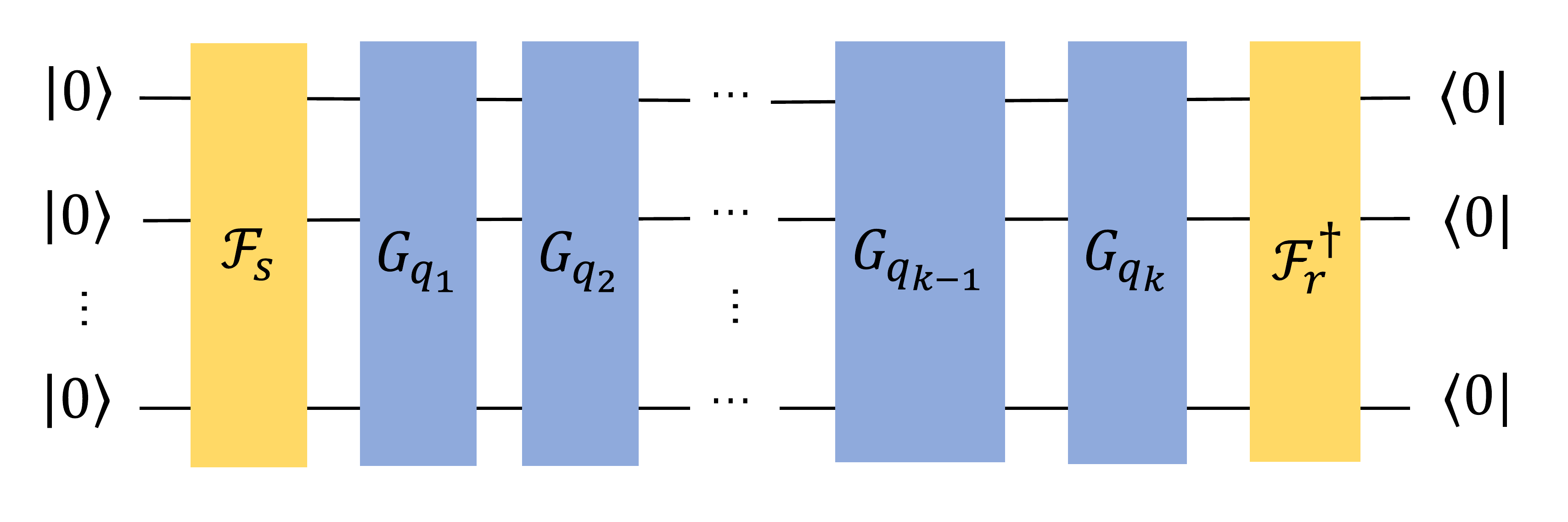}
        \caption{This is a quantum circuit with n-Qubits used to compute Tr\{$\rho_r^\dagger \mathcal{G}_q^k \rho_s\mathcal{G}_q^k$\}. First, we add the prepared state $\rho_r^\dagger$ to the circuit. Next, we apply the operator $\mathcal{G}_q^k$. Then, we add $\rho_s$ to the circuit.}
    \label{fig:4}
\end{figure}

\subsubsection{GST}\label{gst}
In Quantum Process Tomography (QPT), the information required to
reconstruct each gate \(R_{w_q}\) is contained in the measurements of
\(\langle\langle\rho_r|R_{w_q}|\rho_s\rangle\rangle\), and \(R_{w_q}\)
is the PTM of $w_q$ in the Hilbert-Schmidt space.
QPT assumes that the initial state and final measurements are known. In
practice, these states and measurements must be prepared using quantum gates, and these
gates \(\{F_r, F_s^\prime\}\) themselves may have imperfections \cite{greenbaum2015introduction}:
\begin{align}
    \langle\langle \rho_{r}|&=\langle\langle 0|F_{r},\nonumber\\
   |\rho_{s}\rangle\rangle&=F_{s}^{\prime}| 0\rangle\rangle. \nonumber
\end{align}

Indeed, the initial states and final measurements that were prepared using
gates are not directly known and can introduce errors in the estimation
process.
GST aims to characterize the fully
unknown set of gates and states\cite{blume2013robust}.
\begin{equation}
\mathcal{R}=\{|\rho\rangle\rangle,\langle\langle E|, R_{w_1}, \ldots, R_{w_q},\ldots\},q=1,2,\ldots,\alpha^k.  
\end{equation}
GST has similar requirements to QPT: the ability to measure the set of
gates
\(\mathcal{R}=\{|\rho\rangle\rangle, \langle\langle E|, R_{w_1}, \ldots, R_{w_{\alpha^k}}\}\)
in the form of expectation values:
\begin{equation}
p=\Tr\{\rho_r^\dagger w_q(\rho_s)\}=\langle\langle \rho_r|R_{w_q}|\rho_s\rangle\rangle=\Tr\{\rho_r^\dagger w_q\rho_sw_q^\dagger\}.\nonumber   
\end{equation}
To simplify the expression, we use ${{F_{r},F_s^\prime}}$, where $(r,s=1,2,\ldots,d^2)$, to denote the quantum gates used for preparing quantum states and measurements (as shown in Fig~\ref{fig:4}), which are ${G_{q_r}(\theta_r)U_{q_r}}$ and ${U_{q_s}^\dagger G_{q_s}^\dagger(\theta_s)}$, respectively. The density matrices of these prepared quantum states are linearly independent. Please refer to Appendix \ref{Appendix A: Completeness Proof} for details.
By constructing a quantum circuit, it is possible to compute the \(d^2\)
matrix elements of the PTM matrix \(R_{w_q}\). 
We define \((R_{w_q})_{rs}\) as
follows:
\begin{equation}
(R_{w_q})_{rs}=\langle\langle \rho_r|R_{w_q}|\rho_s\rangle\rangle=\langle\langle0|F_rR_{w_q}F_s|0\rangle\rangle.   
\end{equation}
Inserting the completeness state into it yields:
\begin{equation}
\begin{aligned}
p_{rs}&=(R_{w_q})_{rs}=\langle\langle \rho_r|R_{w_q}|\rho_s\rangle\rangle\\&=\langle\langle0|F_rR_{w_q}F_s|0\rangle\rangle\\
&=\sum_{a,b}\langle\langle 0|F_r|a\rangle\rangle\langle\langle a|R_{w_q}|b\rangle\rangle\langle\langle b|F_s|0\rangle\rangle \nonumber\\
&=\sum_{a,b}A_{ra}(R_{w_q})B_{bs}.
\end{aligned}
\end{equation}
It can be easily verified
that:
\begin{equation}
\begin{aligned}
p_{rs}&=(AR_{w_q}B)_{rs},\\
A&=\sum_{r}|r\rangle\rangle\langle\langle 0| F_r,\\
B&=\sum_s F_s|0\rangle\rangle\langle\langle s|.    
\end{aligned}
\end{equation}

Let \(\tilde{R}_{w_q}=AR_{w_q}B\), the identity matrix \(I\) also serves as a mapping, and its PTM 
has the following properties:
\begin{equation}
R_{I}|\rho_s\rangle\rangle=|I(\rho_s)\rangle\rangle=|I\rho_sI^\dagger\rangle\rangle=|\rho_s\rangle\rangle.\nonumber
\end{equation}
It can be observed that the action of \(R_I\) is similar to the identity matrix and its matrix elements can be expressed as follows:
\begin{align}
(R_I)_{rs}&=\langle\langle\rho_r|R_I|\rho_s\rangle\rangle=\Tr\{\rho_r^\dagger I\rho_sI^\dagger\}=\Tr\{\rho_r^\dagger \rho_s\}.    \nonumber
\end{align}
Denoting
\(g_{rs}=(R_I)_{rs}=\langle\langle \rho_r|\rho_s\rangle\rangle=\langle\langle0|F_rF_s|0\rangle\rangle\),
we can insert the completeness state and obtain:
\begin{align}
g_{rs}=\sum_{a,b}\langle\langle 0|F_r|a\rangle\rangle\langle\langle a|b\rangle\rangle\langle\langle b|F_s|0\rangle\rangle=(AB)_{rs}.
\end{align}
For a given combination \(\mathcal{G}^k_q\), let \(\tilde{R}_{I_q}=AB\),
where \(A\) and \(B\) are matrices. The experimental measurement value
\(p_{rs}\) corresponds to the \(rs\) component of the matrix
\(A(R_{w_q})B\), while \(g_{rs}\) corresponds to the \(rs\) component of
the matrix \((AB)\). The quantum channel we reconstruct will differ from the real quantum channel by a similarity transformation:
\begin{equation}
(\tilde{R}_I)^{-1}\tilde{R}_{w_q}=B^{-1}A^{-1}AR_{w_q}B=B^{-1}R_{w_q}B.   \nonumber
\end{equation}
Therefore, we can estimate the trace of $R_{w_q}$:
\begin{equation}
\Tr\{(\tilde{R}_I)^{-1}\tilde{R}_{w_q}\}=\Tr\{B^{-1}R_{w_q}B\}=\Tr\{R_{w_q}\}.
\end{equation}
Based on the calculations mentioned earlier, the value of
\(\Tr\{R_{w_q}\}=|\Tr\{w_q\}|^2\) can be determined. However, this is not
the final result for \(\Tr\{w_q\}\). 

Note: In order to avoid ill-conditioned matrices $R_{I_q}$, it is necessary to perform subspace selection. For detailed analysis, please refer to  Section \ref{Eigenvalue Truncation}.
\subsubsection{Mathematical Processing of Results}\label{MPOR}
To calculate \(\Tr\{w_q\}\), an operation involving taking the square root is required: $|\Tr\{w_q\}|^2=\Tr\{R_{w_q}\}$.
In general, $\Tr\{w_q\}$ can be decomposed into real and imaginary parts:
\begin{equation}
\Tr\{w_q\}=\mathbf{\text{Re}}[\Tr\{w_q\}]+i\cdot \mathbf{\text{Im}}[\Tr\{w_q\}].  \nonumber
\end{equation}
then
\begin{equation}
\Tr\{R_{w_q}\}=(\mathbf{\text{Re}}[\Tr\{w_q\}])^2+(\mathbf{\text{Im}}[\Tr\{w_q\}])^2.\nonumber
\end{equation}
In fact, only the real part needs to be estimated
because the sum of all the imaginary parts of \(\Tr\{w_q\}\) vanishes
after summation.
\begin{equation}
\sum_{q=1}^{\alpha^k}\Tr\{w_q\}=\sum_{q=1}^{\alpha^k}\mathbf{\text{Re}}[\Tr\{w_q\}].
\end{equation}

Next, we show how to estimate the real part of $\Tr\{w_q\}$.
Consider \((w^\prime_q)_{d+1}=(w_q)_d\oplus 1\), where \(w_q^\prime\) is the 
representation of \(\mathcal{G}_q^k\) on the subspace
\(V_q^\prime: span\{|\psi_{q_1}\rangle,|\psi_{q_2}\rangle,\cdot\cdot,|\psi_{q_d}\rangle;|\phi\rangle\}\).
\begin{equation}
w_q' = \begin{pmatrix}
w_q & 0 \\
0 & 1 \\
\end{pmatrix},  \nonumber
\end{equation}
where $|\phi\rangle \notin V_q$ and $|\phi\rangle$ can be prepared through a variational quantum circuit.
It can be seen that \(V_q^\prime\) is also a noninvariant subspace of
\(\mathcal{G}_q^k\). It's obvious that
\begin{equation}
\Tr\{w_q^\prime\}=\Tr\{w_q\}+1,
\label{eq:w}
\end{equation}
and:
\begin{equation}
\begin{aligned}
\Tr\{R_{w^\prime_q}\} &= |\Tr\{w^\prime_q\}|^2 = |\Tr\{w_q\}+1|^2 \\
&= \left|\text{Re}[\Tr\{w_q\}]+i\cdot \text{Im}[\Tr\{w_q\}]+1\right|^2 \\
&= (\text{Re}[\Tr\{w_q\}])^2+2(\text{Re}[\Tr\{w_q\}])\\
&+(\text{Im}[\Tr\{w_q\}])^2+1.  
\end{aligned}
\end{equation}
Recall the relation
\begin{equation}
\begin{aligned}
\Tr\{R_w\}&=|\Tr\{w_q\}|^2\\
&=(\text{Re}[\Tr\{w_q\}])^2+(\text{Im}[\Tr\{w_q\}])^2,\nonumber
\end{aligned}
\end{equation}
we can find that
\begin{align}
\text{Re}[\Tr\{w_q\}]=\frac{1}{2}[\Tr\{R_{w^\prime_q}\}-\Tr\{R_{w_q}\}-1].
\end{align}

The procedure to calculate \(\Tr\{R_{w^\prime_q}\}\) follows a similar
algorithm as for computing \(\Tr\{R_{w_q}\}\), with the distinction that
in this case, \((d+1)^2\) quantum states are required.





\subsection{Algorithm Process of Tomography}
The following is the procedure for calculating $\Tr\{\rho^m\}$ for the quantum state $\rho=\sum_{i=1}^\alpha p_iU_i|0\rangle\langle0|U_i^\dagger$.Algorithm \ref{alg2} serves as the main program for GST, while Algorithm \ref{alg3} functions as a subroutine called multiple times within GST, responsible for iteratively computing $\Tr\{\mathcal{G}_q^k\}$.

\begin{algorithm}
	\renewcommand{\algorithmicrequire}{\textbf{Input:}}

\renewcommand{\algorithmicensure}{\textbf{Output:}}
 \algorithmicrequire{$N$,$m$,$p_i$,$U_i$,$\alpha$,$n$}\\
 \algorithmicensure{$\Tr\{\rho^{m}\}$}
	\caption{}
	\label{alg2}
	\begin{algorithmic}[1]

\For{$\text{iteration}=1$ to $N$}
		\State Set the initial state as $|0\rangle^{\otimes n}$.
        \State Initialise $d\leftarrow 0$
	\State Initialise the subspace $V_q \leftarrow [~] $.
		
		\State // Dimension of the subspace is determined.
            \For{$i=1$ to $m$}
                \State Randomly choose Gate with probability $1/2$ (for $G$ ) or $1/2$ (for $I$ ). \State If it is a $G$ gate, choose $G_i$ with probability $p_i$.
            \EndFor
            \State // Select $\mathcal{G}_q^k$ with probability $\mathcal{P}_q$
            	\For {$i = 1$ to $k$} \State // Here $k$ is the number of $G$ in $G_q^k$. 
	\State Apply gate $U[i]$ to prepare quantum state $|\psi_i\rangle = U[i]|0\rangle^{\otimes n}$. $U[i]$ is the unitary corresponding to $i$th $G$ gate, $G[i] = U[i]G_0 U[i]^{\dagger}$.	
 \State Compute the Gram matrix $g_i$ in the subspace $V \oplus |\psi_i\rangle$.
		\State Compare eigenvalues of Gram matrix $\epsilon_g$ of $g_i$ with threshold $\epsilon$.
  		\State // Perform linear correlation analysis on $|\psi_i\rangle$ states.
		\If{$\epsilon_g \geq \epsilon$}
			\State Update the dimension $d\leftarrow d+1$.		
   \State Update the subspace $V_q \leftarrow V_q\oplus|\psi_i\rangle$.
		\EndIf
        \EndFor
            \State $t_1 \leftarrow$ Subroutine($\mathcal{G}_q^k$, $d$, $V_q$)
            \State $t_2 \leftarrow$ Subroutine($\mathcal{G}_q^k$, $d+1$, $V_q \oplus |\phi\rangle$)
            \State $\Tr\{\mathcal{G}_q^k\}\leftarrow 2^n -d+(t_1-t_2-1)/2$.
            \State // Invoke the subroutine to calculate $\Tr\{\mathcal{G}_q^k\}$.

        \EndFor
            \State Calculate $\Tr\{\rho^{m}\}=\Tr\left\{\left(\frac{1}{2}I-\frac{1}{2}G\right)^m\right\}=\sum_{k=0}^{m}p_kx_k$
            \State  //$p_k = \left(\frac{1}{2}\right)^mC_m^{k}$ and $x_k = (-1)^{k}\Tr\{G^k\}$
	\end{algorithmic}  
\end{algorithm}

\begin{algorithm}
	\renewcommand{\algorithmicrequire}{\textbf{Input:}}
	\renewcommand{\algorithmicensure}{\textbf{Output:}}
 \algorithmicrequire{$\mathcal{G}_q^k$, $d$, $V$}\\
 \algorithmicensure{$\Tr\{R_{w_q}\}$}
	\caption{}
	\label{alg3}
	\begin{algorithmic}[1]
            \For{$\zeta=1$ to $d$}       
            \State $\rho_{q_{\zeta}}=\left|\psi_{q_{\zeta}}\right\rangle\langle\psi_{q_{\zeta}}|=U_{q_{\zeta}}| 0\rangle\langle 0| U_{q_{\zeta}}^{\dagger}$
            
            \EndFor
            
            \For{$\zeta^\prime=1 $ to $d$}   
            \If{$\zeta \neq \zeta^\prime$}
            \State $\rho_{q_{\zeta^\prime }}=\left|\psi_{q_{\zeta^\prime }}\right\rangle\langle\psi_{q_{\zeta^\prime}}|$
            \State $ \quad \quad =G_{\zeta^{\prime}}(\theta_{\zeta^{\prime}})U_{q_\zeta}| 0\rangle\langle 0| U_{q_{\zeta}}^{\dagger} G_{\zeta^{\prime}}^{\dagger}(\theta_{\zeta^{\prime}})$
            \State //Prepare $(d^2-d)$ additional quantum states
            \Else 
            \State Break // Exit the loop
            \EndIf
            \EndFor
        \For{$r=1$ to $d^2$ }
            \For{$s=1$ to $d^2$ }
        \State \Qcircuit @C=0.5cm @R=0.5cm {
        &\lstick{\ket{\psi_s}} &\gate{\mathcal{G}_q^k}&\rstick{\bra{\psi_r}}\qw  }
        \State // $\left(\tilde{R}_{w_{q}}\right)_{r, s}=\left(A R_{w_{q}} B\right)_{r, s}$
        \State \Qcircuit @C=0.5cm @R=0.5cm {
        &\lstick{\ket{\psi_s}} &\gate{I}&\rstick{\bra{\psi_r}}\qw  }
        \State // $\left(\tilde{R}_{I}\right)_{r, s}=\left(A R_{I} B\right)_{r, s}=(A B)_{r, s}$
        \EndFor
        \EndFor
        \State $\Tr\{R_{w_q}\}\leftarrow \Tr\{(\tilde{R}_{I}^{-1}\cdot\tilde{R}_{w_q})\}$.
        
    \end{algorithmic}  
\end{algorithm}



\section{Error Analysis}\label{Error Analysis}

\subsection{Eigenvalue Truncation of the simplest case
\label{Eigenvalue Truncation}}
To prevent the introduction of significant statistical errors during subsequent numerical computations, it becomes imperative to truncate the eigenvalues of the Gram matrix $g$. 
In order to ensure that the eigenvalues of the Gram matrix are all greater than threshold $\epsilon$, we start from $|\psi_2\rangle$ to select the quantum states that constitute the subspace. Here we default to $|\psi_1\rangle$ in the subspace,

Commencing from the iteration $k=2$, the computation of the Gram matrix $g_k$ is carried out. This matrix is associated with four distinct eigenvectors, which can be identified as follows:
\begin{equation}
|\Psi_1\rangle\langle\Psi_1|,|\Psi_1\rangle\langle\Psi_2|,|\Psi_2\rangle\langle\Psi_1|,|\Psi_2\rangle\langle\Psi_2|.
\end{equation}
where $|\Psi_1\rangle = |\psi_1\rangle$ and $|\Psi_2\rangle $ is defined from the Schmidt orthogonalization.
\begin{equation}
|\Psi_2\rangle = \frac{1}{\Delta_2}|\psi_2\rangle - \frac{x_{1,1}}{\Delta_2}|\psi_1\rangle.
\end{equation}
Here $x_{1,1}$ is the overlap  $\langle \psi_1|\psi_2\rangle$ and $|\Delta_2|^2 = 1-|x_{1,1}|^2 $ is a normalization factor.
The eigenvalues are respectively given by
\begin{equation}
1,\frac{|\Delta_2|^2}{1 + |x_{1,1}|^2},\frac{|\Delta_2|^2}{1 + |x_{1,1}|^2},\left|\frac{|\Delta_2|^2}{1 + |x_{1,1}|^2}\right|^2.
\end{equation}
If the smallest eigenvalue $\left|\frac{|\Delta_2|^2}{1 + |x_{1,1}|^2}\right|^2$ is smaller than the threshold $\epsilon$ we pre-set, we add $|\psi_2\rangle$ into the subspace. If not, we discard $|\psi_2\rangle$ and then consider whether $|\psi_3\rangle$ can be added into the subspace.

\subsection{eigenvalue truncation for general cases}
\label{A more general eigenvalue truncation can be applied}
We generalize the discussion in the previous subsection to the case of a higher dimensional subspace. Assume that the current subspace already contains $|\psi_1\rangle, \dots, |\psi_{k-1}\rangle$, and now we need to decide whether we can add $|\psi_k\rangle$ to the subspace.
 $|\psi_k\rangle$ can be written as:
\begin{equation}
|\psi_k\rangle = \sum_i^{k-1} x_{k,i}|\psi_i\rangle + \Delta_k|\Psi_k\rangle,
\end{equation}
where 
\begin{equation}
|\Psi_k\rangle = \frac{1}{\Delta_k}|\psi_k\rangle - \frac{1}{\Delta_k}\sum _{i}^{k-1}x_{k,i}|\psi_i\rangle.
\end{equation}

In this case, the smallest singular value is$|\frac{|\Delta_k|^2}{1 + |x_k|^2}|^2$,where$|x_k|^2 = \Sigma_i^{k-1}|x_{k,i}|^2$.

We must assess the relationship between the smallest singular value and $\epsilon$. If the smallest eigenvalue surpasses $\epsilon$, we should proceed with enlarging the subspace. However, if the smallest singular value is less than $\epsilon$, it is advisable to disregard the state $|\psi_k\rangle$.

\subsection{Eigenvalue Estimation Error}

\label{Eigenvalue Estimation Error}
In the actual process, the measurement of the Gram matrix has statistical errors $\Delta(g)$. This will lead to errors in the calculation of the eigenvalues of the Gram matrix. 
Given that each element of $\Delta(g)$ is subjected to measurement $N$ times, it follows that every element holds an order of magnitude around $O\left(\frac{1}{\sqrt{N}}\right)$. By referencing the Disk Theorem, no eigenvalue surpasses  $O\left(\frac{d^2}{\sqrt{N}}\right)$. Applying the Weyl inequality\cite{cowling1984bandwidth} allows us to deduce that the disparity between the computed minimum eigenvalue of $g$ and the actual minimum eigenvalue doesn't exceed $O\left(\frac{d^2}{\sqrt{N}}\right)$. Employing Hoeffding's inequality, the likelihood of each element within $\Delta(g)$ being less than $\tilde{\epsilon}_g$ amounts to $1-\delta_g < 1 - 2 e^{-2N\tilde{\epsilon}_g^2}$. Consequently, the probability of every element being less than $\tilde{\epsilon}_g$, i,e., the error of the eigenvalue will not exceed $d^2 \Tilde{\epsilon}_g$, can be expressed as $(1-\delta_g)^{d^2} > 1 - d^2 \delta_g = 1 - \tilde{\delta}_g$,
where
\begin{equation}
\tilde{\delta}_g = d^2{\delta_g} \geq 2d^2e^{-2N\tilde{\epsilon}_g^2}.
\end{equation}
This means that the estimation for eigenvalues, to achieve precision $d^2\Tilde{\epsilon}_g$ with a probability of $1-d^2\Tilde{\delta}_g$, requires no more than 
\begin{equation}
N = O\left(\frac{\log{\frac{d^2}{\tilde{\delta}_g}}}{\tilde{\epsilon}_g^2}\right)
\end{equation}
number of measurements 
for each matrix element.

\subsection{The error analysis of $g^{-1}$}
\label{The error analysis of $g^{-1}$}
In the GST process, we need to invert the gram matrix. Next, we consider the effect of the error $\Delta(g)$ on the inverse of $g$.
Using Taylor expansion,
\begin{equation}
\begin{aligned}
(g + \Delta(g))^{-1}&= g^{-1} - g^{-1}\Delta(g)g^{-1} + \\&g^{-1}\Delta(g)g^{-1}\Delta(g)g^{-1} + \cdots,
\end{aligned}
\end{equation}
each matrix term can be bounded as
\begin{equation}
(g^{-1})_{i,j} < \epsilon,\nonumber
\end{equation}
\begin{equation}
\left(g^{-1}\Delta(g)g^{-1}\right)_{i,j} < \frac{d^2 \tilde{\epsilon}_g}{\epsilon^2},\nonumber
\end{equation}
\begin{equation}
\left((g^{-1}\Delta{g})^kg^{-1}\right)_{i,j}< \frac{d^{2k}\tilde{\epsilon}_g^{k}}{\epsilon^{k+1}},\nonumber
\end{equation}
where $i,j = 0,1,\cdots,d-1$. Let  $\tilde{\epsilon}_g = \epsilon_1\epsilon^2$, then :
\begin{equation}
\begin{split}
&\left(g^{-1} - (g+\Delta(g)^{-1}\right)_{i,j} < d^2\epsilon_1\sum_{l=0}^{+\infty}(d^2\epsilon_1\epsilon)^l,\nonumber
\end{split}
\end{equation}
which can be bounded as
\begin{equation}
\left(g^{-1} - (g+\Delta(g))^{-1}\right)_{i,j} < \frac{d^2\epsilon_1}{1 - d^2\epsilon_1\epsilon}.
\end{equation}

\subsection{Sampling Error}
\label{Sampling Error}
In this section, we consider the statistical error when estimating the trace of a quantum gate.
Let $(R_{w_y})_{i,j} = \langle\langle\rho_i|G|\rho_j\rangle\rangle$. The error in computing $\Tr\{g^{-1}R_{w_y}\}$ is given by:
\begin{equation}
\Tr\{\Delta(g^{-1})R_{w_y}\} + \Tr\{g^{-1}\Delta(R_{w_y})\}+ \Tr\{\Delta(g^{-1})\Delta(R_{w_y})\}.
\end{equation}
The first term can be bounded as
\begin{equation}
\Tr\{\Delta(g^{-1})R_{w_y}\} < d^2\Tr\{g^{-1} - (g+\Delta(g))^{-1}\}< \frac{d^4\epsilon_1}{1 - d^2\epsilon_1\epsilon}. \nonumber
\end{equation}
The second term is
\begin{equation}
 \Tr\{g^{-1}\Delta(R_{w_y})\} < \frac{d^2\epsilon_2}{\epsilon},\nonumber
\end{equation}
where we have used
\begin{equation}
\left(\Delta(R_{w_y})\right)_{i,j} < \epsilon_2,\nonumber
\end{equation}
Where $i,j = 0,1,\cdots,d-1$. The last term is
\begin{equation}
\Tr\{\Delta(g^{-1})\Delta(R_{w_y})\} < \frac{d^4\epsilon_1\epsilon_2}{1 - d^2\epsilon_1\epsilon}.\nonumber
\end{equation}
So all the terms must add up to less than
\begin{equation}
\frac{d^4\epsilon_1}{1 - d^2\epsilon_1\epsilon}  + \frac{d^2\epsilon_2}{\epsilon} + \frac{d^4\epsilon_1\epsilon_2}{1 - d^2\epsilon_1\epsilon}. 
\end{equation}

According to Hoeffding's inequality, the probability that one of the terms in $\Delta(R_{w_y})$ is less than $\epsilon_2$ is $1-\delta_2 > 1 - 2 e^{-2N_2\epsilon_2^2}$. Therefore, the probability that each term is less than $\epsilon_2$ is $(1-\delta_2)^{d^2} > 1 - d^2 \delta_2 = 1 - \tilde{\delta}_2$.

Therefore, The matrix elements of a quantum gate do not require more measurements than 
\begin{equation}
N_2 = O\left(\frac{\log{\frac{d^2}{\tilde{\delta}_2}}}{\epsilon_2^2}\right).
\end{equation}

\subsection{Truncation Error}
\label{Truncation Error}
Unlike the error caused by statistical fluctuations, truncation error emerges from the omission of specific quantum states during the subspace construction. We analyze disparities between two quantum circuits: one denoted as $G$, representing the implemented circuit within our setup, and the other denoted as $G'$, derived by substituting a gate layer in $G$. Specifically, $G = G_1 G_2 G_3 \cdots G_n$ constitutes an $n$-layer gate circuit corresponding to states $|\psi_1\rangle, |\psi_2\rangle, \cdots, |\psi_n\rangle$. During subspace construction, certain states like $|\psi_k\rangle$ might be excluded. In the case of $G'$, $|\psi_k\rangle$ is replaced with $|\psi_k^{\prime}\rangle$, effectively eliminating the constituent $|\phi_k\rangle$. To discard $|\phi_k\rangle$, a certain condition must be satisfied.
\begin{equation}
\left|\frac{|\Delta_k|^2}{1+|x_k|^2}\right|^2 < \epsilon.
\end{equation}

In other words, $|\Delta_k|$ can be at most $\epsilon^{1/4}$. Since there is an error in estimating the eigenvalues of $g$, the maximum value of $|\Delta_k|$ is $\epsilon_3 = \left(\epsilon + O(\frac{d^2}{\sqrt{N}})\right)^{1/4}$. Therefore, $|\Tr\{G^{\prime}\} - \Tr\{G\}| \sim O(n\epsilon_3)$, since we discard at most $n$ states $|\phi_j\rangle$. Then $\left||\Tr\{G^{\prime}\}|^2 - |\Tr\{G\}|^2|\right| \sim O(nd\epsilon_3)$.
Let $G$ and $G'$ correspond to $R_{w_y}$ and $R_{w_y}'$, respectively. Then we have$|\Tr\{R_{w_y}^{\prime}\} - \Tr\{R_{w_y}\}| \sim O(dn\epsilon_3)$.


\section{Numerical simulation}\label{results-and-discussion}
\subsection{Quantum state preparation}
\label{1Quantum state preparation}

In the preceding text, we postulated that for the algorithm to be effective, knowledge of the random quantum state's preparation method is imperative. To ensure universality, we express any single-qubit gate through a combination of three fundamental rotation gates. Thus, in this paper, we opt for the ${U}(\theta, \phi, \lambda)$ gate
\begin{equation}
{U}(\theta, \phi, \lambda)=\left(\begin{array}{cc}
\cos (\theta / 2) & -e^{i \lambda} \sin (\theta / 2) \\
e^{i \phi} \sin (\theta / 2) & e^{i \lambda+i \phi} \cos (\theta / 2)
\end{array}\right)
\end{equation}
as the random gate to prepare the random quantum state, allowing us to perform computations using both methods and subsequently compare the outcomes,
where \(\theta\), \(\phi\), and \(\lambda\) are real numbers, and \(i\)
is the imaginary unit.
By encoding this general parameterized gate into a single-qubit quantum gate in the quantum circuit, an $n$-qubit gate can be obtained through tensor product operation: $U^{\otimes n}$.
Different quantum states are prepared by using various quantum gates, and then a random quantum state is simulated by employing a probability-weighted method. Attention, our random gate can naturally extend to the direct product of $n$ distinct single-bit random gates.

\subsection{Processing of the calculation results of the Hadamard Test
}\label{Processing of the calculation results of the Hadamard Test.}

The random gates we need are denoted as $U_{y_s}$, where $s=1,2,\dots,k+1$, $y_s\in\{1,2,3,4\}$. The probability of the random gate $U_{y_s}$ is denoted by $p_{y_s}$. Let $U_{y_s}={U}(\theta_{y_s}, \phi_{y_s}, \lambda_{y_s})$. To simulate the computation of $\Tr\{\rho^{k+1}\}$, the circuit we construct is shown in Fig \ref{fig:5}:

\begin{figure}[t]
        \centering
        \includegraphics[width=1.0\linewidth]{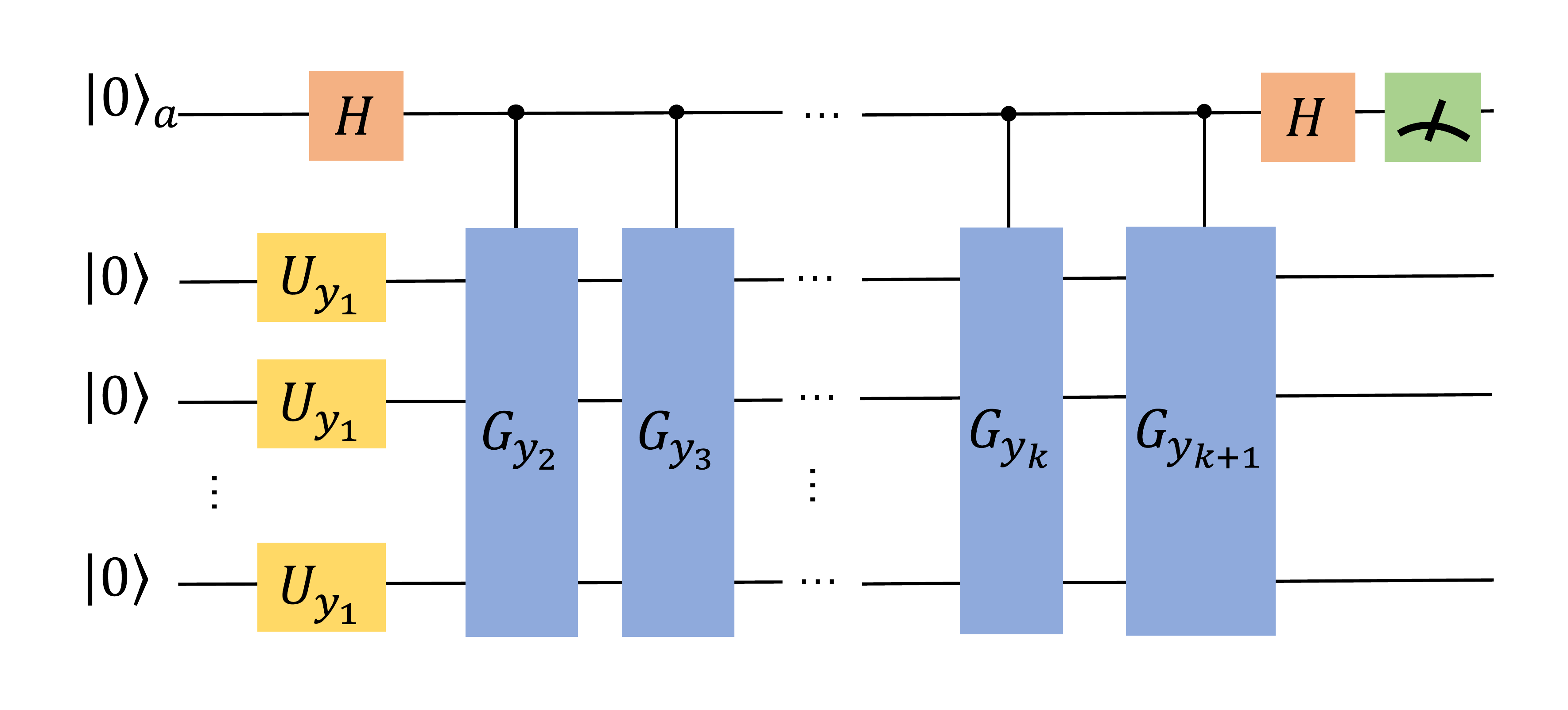}
        \caption{n-qubits Quantum Circuit: The first line represents an ancillary circuit, with the initial state set to $|0\rangle_a$. It then undergoes a Hadamard gate, resulting in a superposition state. The three lines below represent the preparation and measurement of the target quantum states, with initial states all set to $|0\rangle$. After applying the first random gate $U_{y_1}$, the target quantum state $\rho_{y_1} = U_{y_1}|0\rangle\langle0|U_{y_1}^\dagger$ is obtained through tensor product. Next, controlled quantum gates $G_{y_2},G_{y_3} \cdots G_{y_{k+1}}$ are applied, followed by another Hadamard gate on the ancillary circuit. Finally, measurement in the computational basis is performed on the ancillary circuit.}
    \label{fig:5}
\end{figure}

We iterate through $y_s$ to obtain the corresponding $P_y(0)$ for each set of $U_{y_s}$. Summing up all these cases yields the overall $\mathcal{P}(0)$:
\begin{equation}
\mathcal{P}(0)=\sum_{y=1}^{\alpha^{k+1}}\left[\prod_{s=1}^{k+1}p_{y_s}\right]P_{y}(0).  
\end{equation}
Similarly, by following the same procedure we can obtain
\begin{equation}
\mathcal{P}(1)=\sum_{y=1}^{\alpha^{k+1}}\left[\prod_{s=1}^{k+1}p_{y_s}\right]P_{y}(1).    
\end{equation}
With Eq.(\ref{eq12}) and Eq.(\ref{eq13}), we can deduce $\Tr\{G^k\rho\}$, subsequently leading us to the derivation of $\Tr\{\rho^{m+1}\}$.

\subsection{Obtaining the data and exploring potential applications
}\label{Obtaining the data and exploring potential applications}

In this study, the random gates in the circuit are simulated by
selecting different parameters for \({U}(\theta, \phi, \lambda)\). For
generality, the parameter selection is based on random numbers. The
 parameters for the random gates used in this study are shown in Table \ref{tab:perfor1}.
\renewcommand{\arraystretch}{1} 
\begin{table}[b] 
	\fontsize{7.5}{13}\selectfont
	\begin{threeparttable}
		\caption{The values of the parameters}
		\label{tab:perfor1}
		\begin{tabular*}{0.45\textwidth}{@{\extracolsep{\fill}}cccc}
			\hline			
			\multirow{2}{*}& $\theta_{y_s}$ & $\phi_{y_s}$ & $\lambda_{y_s}$ \\			
                \hline					
               
			i=1 & 0.29$\pi$ & 0.07$\pi$ & 0.11$\pi$ \\
			i=2 & 0.46$\pi$ & 0.62$\pi$ & 0.82$\pi$ \\
			i=3 & 0.41$\pi$ & 0.59$\pi$ & 0.53$\pi$ \\
			i=4 & 0.55$\pi$ & 0.31$\pi$ & 0.60$\pi$ \\
                \hline
		\end{tabular*}
           
		\begin{tablenotes}
			\item[*] The parameters of the random gates used to generate the above data.
		\end{tablenotes}
	\end{threeparttable}
\end{table}
The probabilities for the four random gates are 0.1, 0.2, 0.3, and
0.4. Based on the above content, the values of \(\Tr\{\rho^m\}\) obtained
using two different methods are shown in Table \ref{tab:perfor2}.
\renewcommand{\arraystretch}{1} 
\begin{table}[t]   
        \fontsize{7.5}{13}\selectfont
        \renewcommand{\arraystretch}{1} 
        \caption{The numerical value of the parameter}
        \label{tab:perfor2}
        \begin{threeparttable}
            \begin{tabular*}{0.45\textwidth}{@{\extracolsep{\fill}}cccc}
                \hline
                & Hadamard Test & Tomography  \\
                \hline
                $\Tr\{\rho^2\}$ & 0.650& 0.650  \\
                $\Tr\{\rho^3\}$ & 0.486& 0.486 \\
                $\Tr\{\rho^4\}$ & 0.375& 0.375 \\
                \hline
            \end{tabular*}
            \begin{tablenotes}
                \item[*] The computational results obtained using the two algorithms in this article are presented with three decimal places.
            \end{tablenotes}
        \end{threeparttable}
\end{table}
\vspace{3pt}

From the analysis of the data above, we can see that the numerical
values of \(\Tr\{\rho^m\}\) obtained using the HT and GST
are very close. More data considering shot noise can be found in Appendix \ref{Appendix B: More Numerical Simulation}.

However, compared to HT, the GST method doesn't require an additional ancillary qubit, reducing the implementation cost of the quantum circuit. This implies that practical applications could benefit from using fewer physical qubits and controlled gates, which is a critical factor to consider. Reducing the number of controlled gates could lower the error rate of the quantum circuit, thereby enhancing the system's reliability. 

Based on the obtained \(\Tr\{\rho^m\}\), it is possible to calculate
\(\Tr\{\rho \ln\rho\}\), \(\Tr\{e^{\rho t}\}\), \(\Tr\{e^{i\rho t}\}\), etc.
For example \(\Tr\{\rho \ln\rho\}\):
the expansion of \(\rho \ln\rho\) with respect to \(G\) yields the following expression:
\begin{equation}
\begin{aligned}
\rho \ln\rho&\approx-\frac{1}{2}\ln2(I-G)-\frac{1}{2}G+\frac{1}{2}\{(1-\frac{1}{2})G^2+\\&(\frac{1}{2}-\frac{1}{3})G^3+\cdots+(\frac{1}{n-1}-\frac{1}{n})G^n+\frac{1}{n}G^{n+1}\}.
\end{aligned}
\end{equation}

For \(\Tr\{\rho \ln \rho\}\), the theoretical value can be obtained using
the method of matrix multiplication. After expansion, it can be
calculated using the two methods described in this paper.

As shown in Table \ref{tab:perfor3}, it's evident that as the power $m$  increases, the theoretical and experimental values converge, yielding a diminishing relative error. Upon juxtaposing Table \ref{tab:perfor2}, one can discern the remarkable similarity between the outcomes derived from both algorithms. The code can be found in \href{https://github.com/WengchengZhao/Hadamard-Test}{HT} and \href{https://github.com/WengchengZhao/Tomography}{GST}.

\renewcommand{\arraystretch}{1} 
\begin{table}[t]
        \begin{flushleft} 
	\fontsize{7.5}{13}\selectfont
	\begin{threeparttable}
		\caption{Comparison between the calculated $\Tr\{\rho \ln\rho\}$ using the method described in this paper and its theoretical value}
		\label{tab:perfor3}
		\begin{tabular*}{0.45\textwidth}{@{\extracolsep{\fill}}cccccccc}
			\hline
			{ } & $\Tr\{G^m\}$ & T value &M value& RE\\
                \hline
		    $m=2$ & 6.600 & -0.600 & -0.569 & 5.17\% \\
                $m=3$ & 5.914 & -0.600 & -0.543 & 9.50\% \\
                $m=4$ & 6.066 & -0.600 & -0.574 & 4.30\% \\
                $m=5$ & 5.814 & -0.600 & -0.573 & 4.50\% \\
                $m=6$ & 5.830 & -0.600 & -0.582 & 3.00\% \\
                $m=7$ & 5.726 & -0.600 & -0.583 & 2.83\% \\
                $m=8$ & 5.710 & -0.600 & -0.586 & 2.33\% \\
                \hline
		\end{tabular*}
		\begin{tablenotes}
			\item[*] Note: \textbf{T-value} represents the theoretical value of $\Tr\{\rho \ln\rho\}$, and \textbf{M-value} represents the measured value of $\Tr\{\rho \ln\rho\}$.\textbf{RE} represents the relative error.The value of $\Tr\{G^m\}$ is measured using the method described in this article. The measured value of $\Tr(\rho \ln\rho)$ can be calculated using $\Tr\{G^m\}$.
		\end{tablenotes}
	\end{threeparttable}
        \end{flushleft} 
\end{table}

\section{Conclusion}
\label{conclusion}
In this article, we present two algorithms for computing the power function of the density matrix by encoding the quantum state into a quantum channel. The first algorithm is based on the HT. In the absence of noise, this algorithm provides an unbiased estimate of the power function. However, it requires a significant number of control-G gates, which is not favorable for current hardware limitations. Therefore, we propose an alternative algorithm based on GST. The original GST is not scalable, but for our specific problem, we can perform GST only within the non-trivial subspace and extract the necessary information. The advantage of this algorithm lies in significantly reducing the utilization of two-qubit gates. Nevertheless, due to the need to mitigate the impact of shot noise on results, subspace selection introduces minor biases to the outcomes. As an example, we apply both methods to compute the von Neumann entropy of a randomly generated quantum state. It is worth noting that these algorithms can be used not only for computing the power function of the density matrix but also for evaluating nonlinear functions of physical quantities such as $\Tr\{O\rho^m\}$.

\begin{acknowledgments}
\label{acknowledgements}
This work is supported by National Natural Science Foundation of China (Grant No. 12225507, 12088101) and NSAF (Grant No. U1930403).
\end{acknowledgments}

\bibliography{name.bib}

\begin{thebibliography}{47}%
\makeatletter
\providecommand \@ifxundefined [1]{%
 \@ifx{#1\undefined}
}%
\providecommand \@ifnum [1]{%
 \ifnum #1\expandafter \@firstoftwo
 \else \expandafter \@secondoftwo
 \fi
}%
\providecommand \@ifx [1]{%
 \ifx #1\expandafter \@firstoftwo
 \else \expandafter \@secondoftwo
 \fi
}%
\providecommand \natexlab [1]{#1}%
\providecommand \enquote  [1]{``#1''}%
\providecommand \bibnamefont  [1]{#1}%
\providecommand \bibfnamefont [1]{#1}%
\providecommand \citenamefont [1]{#1}%
\providecommand \href@noop [0]{\@secondoftwo}%
\providecommand \href [0]{\begingroup \@sanitize@url \@href}%
\providecommand \@href[1]{\@@startlink{#1}\@@href}%
\providecommand \@@href[1]{\endgroup#1\@@endlink}%
\providecommand \@sanitize@url [0]{\catcode `\\12\catcode `\$12\catcode
  `\&12\catcode `\#12\catcode `\^12\catcode `\_12\catcode `\%12\relax}%
\providecommand \@@startlink[1]{}%
\providecommand \@@endlink[0]{}%
\providecommand \url  [0]{\begingroup\@sanitize@url \@url }%
\providecommand \@url [1]{\endgroup\@href {#1}{\urlprefix }}%
\providecommand \urlprefix  [0]{URL }%
\providecommand \Eprint [0]{\href }%
\providecommand \doibase [0]{http://dx.doi.org/}%
\providecommand \selectlanguage [0]{\@gobble}%
\providecommand \bibinfo  [0]{\@secondoftwo}%
\providecommand \bibfield  [0]{\@secondoftwo}%
\providecommand \translation [1]{[#1]}%
\providecommand \BibitemOpen [0]{}%
\providecommand \bibitemStop [0]{}%
\providecommand \bibitemNoStop [0]{.\EOS\space}%
\providecommand \EOS [0]{\spacefactor3000\relax}%
\providecommand \BibitemShut  [1]{\csname bibitem#1\endcsname}%
\let\auto@bib@innerbib\@empty
\bibitem [{\citenamefont {Swingle}\ \emph {et~al.}(2016)\citenamefont
  {Swingle}, \citenamefont {Bentsen}, \citenamefont {Schleier-Smith},\ and\
  \citenamefont {Hayden}}]{swingle2016measuring}%
  \BibitemOpen
  \bibfield  {author} {\bibinfo {author} {\bibfnamefont {B.}~\bibnamefont
  {Swingle}}, \bibinfo {author} {\bibfnamefont {G.}~\bibnamefont {Bentsen}},
  \bibinfo {author} {\bibfnamefont {M.}~\bibnamefont {Schleier-Smith}}, \ and\
  \bibinfo {author} {\bibfnamefont {P.}~\bibnamefont {Hayden}},\ }\href@noop {}
  {\bibfield  {journal} {\bibinfo  {journal} {Physical Review A}\ }\textbf
  {\bibinfo {volume} {94}},\ \bibinfo {pages} {040302} (\bibinfo {year}
  {2016})}\BibitemShut {NoStop}%
\bibitem [{\citenamefont {Brand{\~a}o}\ \emph {et~al.}(2021)\citenamefont
  {Brand{\~a}o}, \citenamefont {Chemissany}, \citenamefont {Hunter-Jones},
  \citenamefont {Kueng},\ and\ \citenamefont {Preskill}}]{brandao2021models}%
  \BibitemOpen
  \bibfield  {author} {\bibinfo {author} {\bibfnamefont {F.~G.}\ \bibnamefont
  {Brand{\~a}o}}, \bibinfo {author} {\bibfnamefont {W.}~\bibnamefont
  {Chemissany}}, \bibinfo {author} {\bibfnamefont {N.}~\bibnamefont
  {Hunter-Jones}}, \bibinfo {author} {\bibfnamefont {R.}~\bibnamefont {Kueng}},
  \ and\ \bibinfo {author} {\bibfnamefont {J.}~\bibnamefont {Preskill}},\
  }\href@noop {} {\bibfield  {journal} {\bibinfo  {journal} {PRX Quantum}\
  }\textbf {\bibinfo {volume} {2}},\ \bibinfo {pages} {030316} (\bibinfo {year}
  {2021})}\BibitemShut {NoStop}%
\bibitem [{\citenamefont {Hayden}\ and\ \citenamefont
  {Preskill}(2007)}]{hayden2007black}%
  \BibitemOpen
  \bibfield  {author} {\bibinfo {author} {\bibfnamefont {P.}~\bibnamefont
  {Hayden}}\ and\ \bibinfo {author} {\bibfnamefont {J.}~\bibnamefont
  {Preskill}},\ }\href@noop {} {\bibfield  {journal} {\bibinfo  {journal}
  {Journal of high energy physics}\ }\textbf {\bibinfo {volume} {2007}},\
  \bibinfo {pages} {120} (\bibinfo {year} {2007})}\BibitemShut {NoStop}%
\bibitem [{\citenamefont {Kudler-Flam}(2021)}]{kudler2021relative}%
  \BibitemOpen
  \bibfield  {author} {\bibinfo {author} {\bibfnamefont {J.}~\bibnamefont
  {Kudler-Flam}},\ }\href@noop {} {\bibfield  {journal} {\bibinfo  {journal}
  {Physical Review Letters}\ }\textbf {\bibinfo {volume} {126}},\ \bibinfo
  {pages} {171603} (\bibinfo {year} {2021})}\BibitemShut {NoStop}%
\bibitem [{\citenamefont {Holmes}\ \emph {et~al.}(2023)\citenamefont {Holmes},
  \citenamefont {Coble}, \citenamefont {Sornborger},\ and\ \citenamefont
  {Suba{\c{s}}{\i}}}]{holmes2023nonlinear}%
  \BibitemOpen
  \bibfield  {author} {\bibinfo {author} {\bibfnamefont {Z.}~\bibnamefont
  {Holmes}}, \bibinfo {author} {\bibfnamefont {N.~J.}\ \bibnamefont {Coble}},
  \bibinfo {author} {\bibfnamefont {A.~T.}\ \bibnamefont {Sornborger}}, \ and\
  \bibinfo {author} {\bibfnamefont {Y.}~\bibnamefont {Suba{\c{s}}{\i}}},\
  }\href@noop {} {\bibfield  {journal} {\bibinfo  {journal} {Physical Review
  Research}\ }\textbf {\bibinfo {volume} {5}},\ \bibinfo {pages} {013105}
  (\bibinfo {year} {2023})}\BibitemShut {NoStop}%
\bibitem [{\citenamefont {Subramanian}\ and\ \citenamefont
  {Hsieh}(2021)}]{subramanian2021quantum}%
  \BibitemOpen
  \bibfield  {author} {\bibinfo {author} {\bibfnamefont {S.}~\bibnamefont
  {Subramanian}}\ and\ \bibinfo {author} {\bibfnamefont {M.-H.}\ \bibnamefont
  {Hsieh}},\ }\href@noop {} {\bibfield  {journal} {\bibinfo  {journal}
  {Physical review A}\ }\textbf {\bibinfo {volume} {104}},\ \bibinfo {pages}
  {022428} (\bibinfo {year} {2021})}\BibitemShut {NoStop}%
\bibitem [{\citenamefont {Zyczkowski}\ and\ \citenamefont
  {Bengtsson}(2006)}]{zyczkowski2006introduction}%
  \BibitemOpen
  \bibfield  {author} {\bibinfo {author} {\bibfnamefont {K.}~\bibnamefont
  {Zyczkowski}}\ and\ \bibinfo {author} {\bibfnamefont {I.}~\bibnamefont
  {Bengtsson}},\ }\href@noop {} {\bibfield  {journal} {\bibinfo  {journal}
  {arXiv preprint quant-ph/0606228}\ } (\bibinfo {year} {2006})}\BibitemShut
  {NoStop}%
\bibitem [{\citenamefont {Rath}\ \emph {et~al.}(2021)\citenamefont {Rath},
  \citenamefont {Branciard}, \citenamefont {Minguzzi},\ and\ \citenamefont
  {Vermersch}}]{rath2021quantum}%
  \BibitemOpen
  \bibfield  {author} {\bibinfo {author} {\bibfnamefont {A.}~\bibnamefont
  {Rath}}, \bibinfo {author} {\bibfnamefont {C.}~\bibnamefont {Branciard}},
  \bibinfo {author} {\bibfnamefont {A.}~\bibnamefont {Minguzzi}}, \ and\
  \bibinfo {author} {\bibfnamefont {B.}~\bibnamefont {Vermersch}},\ }\href@noop
  {} {\bibfield  {journal} {\bibinfo  {journal} {Physical Review Letters}\
  }\textbf {\bibinfo {volume} {127}},\ \bibinfo {pages} {260501} (\bibinfo
  {year} {2021})}\BibitemShut {NoStop}%
\bibitem [{\citenamefont {Jozsa}(1994)}]{jozsa1994fidelity}%
  \BibitemOpen
  \bibfield  {author} {\bibinfo {author} {\bibfnamefont {R.}~\bibnamefont
  {Jozsa}},\ }\href@noop {} {\bibfield  {journal} {\bibinfo  {journal} {Journal
  of modern optics}\ }\textbf {\bibinfo {volume} {41}},\ \bibinfo {pages}
  {2315} (\bibinfo {year} {1994})}\BibitemShut {NoStop}%
\bibitem [{\citenamefont {Koczor}(2021)}]{koczor2021exponential}%
  \BibitemOpen
  \bibfield  {author} {\bibinfo {author} {\bibfnamefont {B.}~\bibnamefont
  {Koczor}},\ }\href@noop {} {\bibfield  {journal} {\bibinfo  {journal}
  {Physical Review X}\ }\textbf {\bibinfo {volume} {11}},\ \bibinfo {pages}
  {031057} (\bibinfo {year} {2021})}\BibitemShut {NoStop}%
\bibitem [{\citenamefont {Wang}\ \emph {et~al.}(2022)\citenamefont {Wang},
  \citenamefont {Guan}, \citenamefont {Liu}, \citenamefont {Zhang},\ and\
  \citenamefont {Ying}}]{wang2022new}%
  \BibitemOpen
  \bibfield  {author} {\bibinfo {author} {\bibfnamefont {Q.}~\bibnamefont
  {Wang}}, \bibinfo {author} {\bibfnamefont {J.}~\bibnamefont {Guan}}, \bibinfo
  {author} {\bibfnamefont {J.}~\bibnamefont {Liu}}, \bibinfo {author}
  {\bibfnamefont {Z.}~\bibnamefont {Zhang}}, \ and\ \bibinfo {author}
  {\bibfnamefont {M.}~\bibnamefont {Ying}},\ }\href@noop {} {\bibfield
  {journal} {\bibinfo  {journal} {arXiv preprint arXiv:2203.13522}\ } (\bibinfo
  {year} {2022})}\BibitemShut {NoStop}%
\bibitem [{\citenamefont {Kandala}\ \emph {et~al.}(2017)\citenamefont
  {Kandala}, \citenamefont {Mezzacapo}, \citenamefont {Temme}, \citenamefont
  {Takita}, \citenamefont {Brink}, \citenamefont {Chow},\ and\ \citenamefont
  {Gambetta}}]{kandala2017hardware}%
  \BibitemOpen
  \bibfield  {author} {\bibinfo {author} {\bibfnamefont {A.}~\bibnamefont
  {Kandala}}, \bibinfo {author} {\bibfnamefont {A.}~\bibnamefont {Mezzacapo}},
  \bibinfo {author} {\bibfnamefont {K.}~\bibnamefont {Temme}}, \bibinfo
  {author} {\bibfnamefont {M.}~\bibnamefont {Takita}}, \bibinfo {author}
  {\bibfnamefont {M.}~\bibnamefont {Brink}}, \bibinfo {author} {\bibfnamefont
  {J.~M.}\ \bibnamefont {Chow}}, \ and\ \bibinfo {author} {\bibfnamefont
  {J.~M.}\ \bibnamefont {Gambetta}},\ }\href@noop {} {\bibfield  {journal}
  {\bibinfo  {journal} {nature}\ }\textbf {\bibinfo {volume} {549}},\ \bibinfo
  {pages} {242} (\bibinfo {year} {2017})}\BibitemShut {NoStop}%
\bibitem [{\citenamefont {Carteret}(2005)}]{carteret2005noiseless}%
  \BibitemOpen
  \bibfield  {author} {\bibinfo {author} {\bibfnamefont {H.~A.}\ \bibnamefont
  {Carteret}},\ }\href@noop {} {\bibfield  {journal} {\bibinfo  {journal}
  {Physical review letters}\ }\textbf {\bibinfo {volume} {94}},\ \bibinfo
  {pages} {040502} (\bibinfo {year} {2005})}\BibitemShut {NoStop}%
\bibitem [{\citenamefont {Lubasch}\ \emph {et~al.}(2020)\citenamefont
  {Lubasch}, \citenamefont {Joo}, \citenamefont {Moinier}, \citenamefont
  {Kiffner},\ and\ \citenamefont {Jaksch}}]{lubasch2020variational}%
  \BibitemOpen
  \bibfield  {author} {\bibinfo {author} {\bibfnamefont {M.}~\bibnamefont
  {Lubasch}}, \bibinfo {author} {\bibfnamefont {J.}~\bibnamefont {Joo}},
  \bibinfo {author} {\bibfnamefont {P.}~\bibnamefont {Moinier}}, \bibinfo
  {author} {\bibfnamefont {M.}~\bibnamefont {Kiffner}}, \ and\ \bibinfo
  {author} {\bibfnamefont {D.}~\bibnamefont {Jaksch}},\ }\href@noop {}
  {\bibfield  {journal} {\bibinfo  {journal} {Physical Review A}\ }\textbf
  {\bibinfo {volume} {101}},\ \bibinfo {pages} {010301} (\bibinfo {year}
  {2020})}\BibitemShut {NoStop}%
\bibitem [{\citenamefont {Georgeot}\ and\ \citenamefont
  {Shepelyansky}(2001)}]{georgeot2001exponential}%
  \BibitemOpen
  \bibfield  {author} {\bibinfo {author} {\bibfnamefont {B.}~\bibnamefont
  {Georgeot}}\ and\ \bibinfo {author} {\bibfnamefont {D.~L.}\ \bibnamefont
  {Shepelyansky}},\ }\href@noop {} {\bibfield  {journal} {\bibinfo  {journal}
  {Physical Review Letters}\ }\textbf {\bibinfo {volume} {86}},\ \bibinfo
  {pages} {2890} (\bibinfo {year} {2001})}\BibitemShut {NoStop}%
\bibitem [{\citenamefont {Elben}\ \emph
  {et~al.}(2020{\natexlab{a}})\citenamefont {Elben}, \citenamefont {Yu},
  \citenamefont {Zhu}, \citenamefont {Hafezi}, \citenamefont {Pollmann},
  \citenamefont {Zoller},\ and\ \citenamefont {Vermersch}}]{elben2020many}%
  \BibitemOpen
  \bibfield  {author} {\bibinfo {author} {\bibfnamefont {A.}~\bibnamefont
  {Elben}}, \bibinfo {author} {\bibfnamefont {J.}~\bibnamefont {Yu}}, \bibinfo
  {author} {\bibfnamefont {G.}~\bibnamefont {Zhu}}, \bibinfo {author}
  {\bibfnamefont {M.}~\bibnamefont {Hafezi}}, \bibinfo {author} {\bibfnamefont
  {F.}~\bibnamefont {Pollmann}}, \bibinfo {author} {\bibfnamefont
  {P.}~\bibnamefont {Zoller}}, \ and\ \bibinfo {author} {\bibfnamefont
  {B.}~\bibnamefont {Vermersch}},\ }\href@noop {} {\bibfield  {journal}
  {\bibinfo  {journal} {Science advances}\ }\textbf {\bibinfo {volume} {6}},\
  \bibinfo {pages} {eaaz3666} (\bibinfo {year}
  {2020}{\natexlab{a}})}\BibitemShut {NoStop}%
\bibitem [{\citenamefont {Braum{\"u}ller}\ \emph {et~al.}(2022)\citenamefont
  {Braum{\"u}ller}, \citenamefont {Karamlou}, \citenamefont {Yanay},
  \citenamefont {Kannan}, \citenamefont {Kim}, \citenamefont {Kjaergaard},
  \citenamefont {Melville}, \citenamefont {Niedzielski}, \citenamefont {Sung},
  \citenamefont {Veps{\"a}l{\"a}inen} \emph {et~al.}}]{braumuller2022probing}%
  \BibitemOpen
  \bibfield  {author} {\bibinfo {author} {\bibfnamefont {J.}~\bibnamefont
  {Braum{\"u}ller}}, \bibinfo {author} {\bibfnamefont {A.~H.}\ \bibnamefont
  {Karamlou}}, \bibinfo {author} {\bibfnamefont {Y.}~\bibnamefont {Yanay}},
  \bibinfo {author} {\bibfnamefont {B.}~\bibnamefont {Kannan}}, \bibinfo
  {author} {\bibfnamefont {D.}~\bibnamefont {Kim}}, \bibinfo {author}
  {\bibfnamefont {M.}~\bibnamefont {Kjaergaard}}, \bibinfo {author}
  {\bibfnamefont {A.}~\bibnamefont {Melville}}, \bibinfo {author}
  {\bibfnamefont {B.~M.}\ \bibnamefont {Niedzielski}}, \bibinfo {author}
  {\bibfnamefont {Y.}~\bibnamefont {Sung}}, \bibinfo {author} {\bibfnamefont
  {A.}~\bibnamefont {Veps{\"a}l{\"a}inen}},  \emph {et~al.},\ }\href@noop {}
  {\bibfield  {journal} {\bibinfo  {journal} {Nature Physics}\ }\textbf
  {\bibinfo {volume} {18}},\ \bibinfo {pages} {172} (\bibinfo {year}
  {2022})}\BibitemShut {NoStop}%
\bibitem [{\citenamefont {of~Sciences~Engineering}\ \emph
  {et~al.}(2019)\citenamefont {of~Sciences~Engineering}, \citenamefont
  {Medicine} \emph {et~al.}}]{national2019quantum}%
  \BibitemOpen
  \bibfield  {author} {\bibinfo {author} {\bibfnamefont {N.~A.}\ \bibnamefont
  {of~Sciences~Engineering}}, \bibinfo {author} {\bibnamefont {Medicine}},
  \emph {et~al.},\ }\href@noop {} {\  (\bibinfo {year} {2019})}\BibitemShut
  {NoStop}%
\bibitem [{\citenamefont {Preskill}(2018)}]{preskill2018quantum}%
  \BibitemOpen
  \bibfield  {author} {\bibinfo {author} {\bibfnamefont {J.}~\bibnamefont
  {Preskill}},\ }\href@noop {} {\bibfield  {journal} {\bibinfo  {journal}
  {Quantum}\ }\textbf {\bibinfo {volume} {2}},\ \bibinfo {pages} {79} (\bibinfo
  {year} {2018})}\BibitemShut {NoStop}%
\bibitem [{\citenamefont {Lau}\ \emph {et~al.}(2022)\citenamefont {Lau},
  \citenamefont {Lim}, \citenamefont {Shrotriya},\ and\ \citenamefont
  {Kwek}}]{lau2022nisq}%
  \BibitemOpen
  \bibfield  {author} {\bibinfo {author} {\bibfnamefont {J.~W.~Z.}\
  \bibnamefont {Lau}}, \bibinfo {author} {\bibfnamefont {K.~H.}\ \bibnamefont
  {Lim}}, \bibinfo {author} {\bibfnamefont {H.}~\bibnamefont {Shrotriya}}, \
  and\ \bibinfo {author} {\bibfnamefont {L.~C.}\ \bibnamefont {Kwek}},\
  }\href@noop {} {\bibfield  {journal} {\bibinfo  {journal} {AAPPS Bulletin}\
  }\textbf {\bibinfo {volume} {32}},\ \bibinfo {pages} {27} (\bibinfo {year}
  {2022})}\BibitemShut {NoStop}%
\bibitem [{\citenamefont {Ding}\ and\ \citenamefont
  {Chong}(2022)}]{ding2022quantum}%
  \BibitemOpen
  \bibfield  {author} {\bibinfo {author} {\bibfnamefont {Y.}~\bibnamefont
  {Ding}}\ and\ \bibinfo {author} {\bibfnamefont {F.~T.}\ \bibnamefont
  {Chong}},\ }\href@noop {} {\emph {\bibinfo {title} {Quantum computer systems:
  Research for noisy intermediate-scale quantum computers}}}\ (\bibinfo
  {publisher} {Springer Nature},\ \bibinfo {year} {2022})\BibitemShut {NoStop}%
\bibitem [{\citenamefont {Zhou}\ and\ \citenamefont
  {Liu}(2022)}]{zhou2022hybrid}%
  \BibitemOpen
  \bibfield  {author} {\bibinfo {author} {\bibfnamefont {Y.}~\bibnamefont
  {Zhou}}\ and\ \bibinfo {author} {\bibfnamefont {Z.}~\bibnamefont {Liu}},\
  }\href@noop {} {\bibfield  {journal} {\bibinfo  {journal} {arXiv preprint
  arXiv:2208.08416}\ } (\bibinfo {year} {2022})}\BibitemShut {NoStop}%
\bibitem [{\citenamefont {Bovino}\ \emph {et~al.}(2005)\citenamefont {Bovino},
  \citenamefont {Castagnoli}, \citenamefont {Ekert}, \citenamefont {Horodecki},
  \citenamefont {Alves},\ and\ \citenamefont {Sergienko}}]{bovino2005direct}%
  \BibitemOpen
  \bibfield  {author} {\bibinfo {author} {\bibfnamefont {F.~A.}\ \bibnamefont
  {Bovino}}, \bibinfo {author} {\bibfnamefont {G.}~\bibnamefont {Castagnoli}},
  \bibinfo {author} {\bibfnamefont {A.}~\bibnamefont {Ekert}}, \bibinfo
  {author} {\bibfnamefont {P.}~\bibnamefont {Horodecki}}, \bibinfo {author}
  {\bibfnamefont {C.~M.}\ \bibnamefont {Alves}}, \ and\ \bibinfo {author}
  {\bibfnamefont {A.~V.}\ \bibnamefont {Sergienko}},\ }\href@noop {} {\bibfield
   {journal} {\bibinfo  {journal} {Physical review letters}\ }\textbf {\bibinfo
  {volume} {95}},\ \bibinfo {pages} {240407} (\bibinfo {year}
  {2005})}\BibitemShut {NoStop}%
\bibitem [{\citenamefont {Horodecki}(2003)}]{horodecki2003measuring}%
  \BibitemOpen
  \bibfield  {author} {\bibinfo {author} {\bibfnamefont {P.}~\bibnamefont
  {Horodecki}},\ }\href@noop {} {\bibfield  {journal} {\bibinfo  {journal}
  {Physical review letters}\ }\textbf {\bibinfo {volume} {90}},\ \bibinfo
  {pages} {167901} (\bibinfo {year} {2003})}\BibitemShut {NoStop}%
\bibitem [{\citenamefont {Ekert}\ \emph {et~al.}(2002)\citenamefont {Ekert},
  \citenamefont {Alves}, \citenamefont {Oi}, \citenamefont {Horodecki},
  \citenamefont {Horodecki},\ and\ \citenamefont {Kwek}}]{ekert2002direct}%
  \BibitemOpen
  \bibfield  {author} {\bibinfo {author} {\bibfnamefont {A.~K.}\ \bibnamefont
  {Ekert}}, \bibinfo {author} {\bibfnamefont {C.~M.}\ \bibnamefont {Alves}},
  \bibinfo {author} {\bibfnamefont {D.~K.}\ \bibnamefont {Oi}}, \bibinfo
  {author} {\bibfnamefont {M.}~\bibnamefont {Horodecki}}, \bibinfo {author}
  {\bibfnamefont {P.}~\bibnamefont {Horodecki}}, \ and\ \bibinfo {author}
  {\bibfnamefont {L.~C.}\ \bibnamefont {Kwek}},\ }\href@noop {} {\bibfield
  {journal} {\bibinfo  {journal} {Physical review letters}\ }\textbf {\bibinfo
  {volume} {88}},\ \bibinfo {pages} {217901} (\bibinfo {year}
  {2002})}\BibitemShut {NoStop}%
\bibitem [{\citenamefont {Sack}\ \emph {et~al.}(2022)\citenamefont {Sack},
  \citenamefont {Medina}, \citenamefont {Michailidis}, \citenamefont {Kueng},\
  and\ \citenamefont {Serbyn}}]{sack2022avoiding}%
  \BibitemOpen
  \bibfield  {author} {\bibinfo {author} {\bibfnamefont {S.~H.}\ \bibnamefont
  {Sack}}, \bibinfo {author} {\bibfnamefont {R.~A.}\ \bibnamefont {Medina}},
  \bibinfo {author} {\bibfnamefont {A.~A.}\ \bibnamefont {Michailidis}},
  \bibinfo {author} {\bibfnamefont {R.}~\bibnamefont {Kueng}}, \ and\ \bibinfo
  {author} {\bibfnamefont {M.}~\bibnamefont {Serbyn}},\ }\href@noop {}
  {\bibfield  {journal} {\bibinfo  {journal} {PRX Quantum}\ }\textbf {\bibinfo
  {volume} {3}},\ \bibinfo {pages} {020365} (\bibinfo {year}
  {2022})}\BibitemShut {NoStop}%
\bibitem [{\citenamefont {Zhang}\ \emph {et~al.}(2021)\citenamefont {Zhang},
  \citenamefont {Sun}, \citenamefont {Fang}, \citenamefont {Zhang},
  \citenamefont {Yuan},\ and\ \citenamefont {Lu}}]{zhang2021experimental}%
  \BibitemOpen
  \bibfield  {author} {\bibinfo {author} {\bibfnamefont {T.}~\bibnamefont
  {Zhang}}, \bibinfo {author} {\bibfnamefont {J.}~\bibnamefont {Sun}}, \bibinfo
  {author} {\bibfnamefont {X.-X.}\ \bibnamefont {Fang}}, \bibinfo {author}
  {\bibfnamefont {X.-M.}\ \bibnamefont {Zhang}}, \bibinfo {author}
  {\bibfnamefont {X.}~\bibnamefont {Yuan}}, \ and\ \bibinfo {author}
  {\bibfnamefont {H.}~\bibnamefont {Lu}},\ }\href@noop {} {\bibfield  {journal}
  {\bibinfo  {journal} {Physical Review Letters}\ }\textbf {\bibinfo {volume}
  {127}},\ \bibinfo {pages} {200501} (\bibinfo {year} {2021})}\BibitemShut
  {NoStop}%
\bibitem [{\citenamefont {Seif}\ \emph {et~al.}(2023)\citenamefont {Seif},
  \citenamefont {Cian}, \citenamefont {Zhou}, \citenamefont {Chen},\ and\
  \citenamefont {Jiang}}]{seif2023shadow}%
  \BibitemOpen
  \bibfield  {author} {\bibinfo {author} {\bibfnamefont {A.}~\bibnamefont
  {Seif}}, \bibinfo {author} {\bibfnamefont {Z.-P.}\ \bibnamefont {Cian}},
  \bibinfo {author} {\bibfnamefont {S.}~\bibnamefont {Zhou}}, \bibinfo {author}
  {\bibfnamefont {S.}~\bibnamefont {Chen}}, \ and\ \bibinfo {author}
  {\bibfnamefont {L.}~\bibnamefont {Jiang}},\ }\href@noop {} {\bibfield
  {journal} {\bibinfo  {journal} {PRX Quantum}\ }\textbf {\bibinfo {volume}
  {4}},\ \bibinfo {pages} {010303} (\bibinfo {year} {2023})}\BibitemShut
  {NoStop}%
\bibitem [{\citenamefont {Elben}\ \emph {et~al.}(2023)\citenamefont {Elben},
  \citenamefont {Flammia}, \citenamefont {Huang}, \citenamefont {Kueng},
  \citenamefont {Preskill}, \citenamefont {Vermersch},\ and\ \citenamefont
  {Zoller}}]{elben2023randomized}%
  \BibitemOpen
  \bibfield  {author} {\bibinfo {author} {\bibfnamefont {A.}~\bibnamefont
  {Elben}}, \bibinfo {author} {\bibfnamefont {S.~T.}\ \bibnamefont {Flammia}},
  \bibinfo {author} {\bibfnamefont {H.-Y.}\ \bibnamefont {Huang}}, \bibinfo
  {author} {\bibfnamefont {R.}~\bibnamefont {Kueng}}, \bibinfo {author}
  {\bibfnamefont {J.}~\bibnamefont {Preskill}}, \bibinfo {author}
  {\bibfnamefont {B.}~\bibnamefont {Vermersch}}, \ and\ \bibinfo {author}
  {\bibfnamefont {P.}~\bibnamefont {Zoller}},\ }\href@noop {} {\bibfield
  {journal} {\bibinfo  {journal} {Nature Reviews Physics}\ }\textbf {\bibinfo
  {volume} {5}},\ \bibinfo {pages} {9} (\bibinfo {year} {2023})}\BibitemShut
  {NoStop}%
\bibitem [{\citenamefont {Brydges}\ \emph {et~al.}(2019)\citenamefont
  {Brydges}, \citenamefont {Elben}, \citenamefont {Jurcevic}, \citenamefont
  {Vermersch}, \citenamefont {Maier}, \citenamefont {Lanyon}, \citenamefont
  {Zoller}, \citenamefont {Blatt},\ and\ \citenamefont
  {Roos}}]{brydges2019probing}%
  \BibitemOpen
  \bibfield  {author} {\bibinfo {author} {\bibfnamefont {T.}~\bibnamefont
  {Brydges}}, \bibinfo {author} {\bibfnamefont {A.}~\bibnamefont {Elben}},
  \bibinfo {author} {\bibfnamefont {P.}~\bibnamefont {Jurcevic}}, \bibinfo
  {author} {\bibfnamefont {B.}~\bibnamefont {Vermersch}}, \bibinfo {author}
  {\bibfnamefont {C.}~\bibnamefont {Maier}}, \bibinfo {author} {\bibfnamefont
  {B.~P.}\ \bibnamefont {Lanyon}}, \bibinfo {author} {\bibfnamefont
  {P.}~\bibnamefont {Zoller}}, \bibinfo {author} {\bibfnamefont
  {R.}~\bibnamefont {Blatt}}, \ and\ \bibinfo {author} {\bibfnamefont {C.~F.}\
  \bibnamefont {Roos}},\ }\href@noop {} {\bibfield  {journal} {\bibinfo
  {journal} {Science}\ }\textbf {\bibinfo {volume} {364}},\ \bibinfo {pages}
  {260} (\bibinfo {year} {2019})}\BibitemShut {NoStop}%
\bibitem [{\citenamefont {Elben}\ \emph {et~al.}(2019)\citenamefont {Elben},
  \citenamefont {Vermersch}, \citenamefont {Roos},\ and\ \citenamefont
  {Zoller}}]{elben2019statistical}%
  \BibitemOpen
  \bibfield  {author} {\bibinfo {author} {\bibfnamefont {A.}~\bibnamefont
  {Elben}}, \bibinfo {author} {\bibfnamefont {B.}~\bibnamefont {Vermersch}},
  \bibinfo {author} {\bibfnamefont {C.~F.}\ \bibnamefont {Roos}}, \ and\
  \bibinfo {author} {\bibfnamefont {P.}~\bibnamefont {Zoller}},\ }\href@noop {}
  {\bibfield  {journal} {\bibinfo  {journal} {Physical Review A}\ }\textbf
  {\bibinfo {volume} {99}},\ \bibinfo {pages} {052323} (\bibinfo {year}
  {2019})}\BibitemShut {NoStop}%
\bibitem [{\citenamefont {Elben}\ \emph
  {et~al.}(2020{\natexlab{b}})\citenamefont {Elben}, \citenamefont {Kueng},
  \citenamefont {Huang}, \citenamefont {van Bijnen}, \citenamefont {Kokail},
  \citenamefont {Dalmonte}, \citenamefont {Calabrese}, \citenamefont {Kraus},
  \citenamefont {Preskill}, \citenamefont {Zoller} \emph
  {et~al.}}]{elben2020mixed}%
  \BibitemOpen
  \bibfield  {author} {\bibinfo {author} {\bibfnamefont {A.}~\bibnamefont
  {Elben}}, \bibinfo {author} {\bibfnamefont {R.}~\bibnamefont {Kueng}},
  \bibinfo {author} {\bibfnamefont {H.-Y.~R.}\ \bibnamefont {Huang}}, \bibinfo
  {author} {\bibfnamefont {R.}~\bibnamefont {van Bijnen}}, \bibinfo {author}
  {\bibfnamefont {C.}~\bibnamefont {Kokail}}, \bibinfo {author} {\bibfnamefont
  {M.}~\bibnamefont {Dalmonte}}, \bibinfo {author} {\bibfnamefont
  {P.}~\bibnamefont {Calabrese}}, \bibinfo {author} {\bibfnamefont
  {B.}~\bibnamefont {Kraus}}, \bibinfo {author} {\bibfnamefont
  {J.}~\bibnamefont {Preskill}}, \bibinfo {author} {\bibfnamefont
  {P.}~\bibnamefont {Zoller}},  \emph {et~al.},\ }\href@noop {} {\bibfield
  {journal} {\bibinfo  {journal} {Physical Review Letters}\ }\textbf {\bibinfo
  {volume} {125}},\ \bibinfo {pages} {200501} (\bibinfo {year}
  {2020}{\natexlab{b}})}\BibitemShut {NoStop}%
\bibitem [{\citenamefont {O'Donnell}\ and\ \citenamefont
  {Wright}(2016)}]{o2016efficient}%
  \BibitemOpen
  \bibfield  {author} {\bibinfo {author} {\bibfnamefont {R.}~\bibnamefont
  {O'Donnell}}\ and\ \bibinfo {author} {\bibfnamefont {J.}~\bibnamefont
  {Wright}},\ }in\ \href@noop {} {\emph {\bibinfo {booktitle} {Proceedings of
  the forty-eighth annual ACM symposium on Theory of Computing}}}\ (\bibinfo
  {year} {2016})\ pp.\ \bibinfo {pages} {899--912}\BibitemShut {NoStop}%
\bibitem [{\citenamefont {D'Ariano}\ \emph {et~al.}(2003)\citenamefont
  {D'Ariano}, \citenamefont {Paris},\ and\ \citenamefont
  {Sacchi}}]{d2003quantum}%
  \BibitemOpen
  \bibfield  {author} {\bibinfo {author} {\bibfnamefont {G.~M.}\ \bibnamefont
  {D'Ariano}}, \bibinfo {author} {\bibfnamefont {M.~G.}\ \bibnamefont {Paris}},
  \ and\ \bibinfo {author} {\bibfnamefont {M.~F.}\ \bibnamefont {Sacchi}},\
  }\href@noop {} {\bibfield  {journal} {\bibinfo  {journal} {Advances in
  imaging and electron physics}\ }\textbf {\bibinfo {volume} {128}},\ \bibinfo
  {pages} {206} (\bibinfo {year} {2003})}\BibitemShut {NoStop}%
\bibitem [{\citenamefont {Yang}\ and\ \citenamefont
  {Li}(2021)}]{yang2021perturbative}%
  \BibitemOpen
  \bibfield  {author} {\bibinfo {author} {\bibfnamefont {R.}~\bibnamefont
  {Yang}}\ and\ \bibinfo {author} {\bibfnamefont {Y.}~\bibnamefont {Li}},\
  }\href@noop {} {\bibfield  {journal} {\bibinfo  {journal} {Physical Review
  A}\ }\textbf {\bibinfo {volume} {103}},\ \bibinfo {pages} {032421} (\bibinfo
  {year} {2021})}\BibitemShut {NoStop}%
\bibitem [{\citenamefont {Wu}\ \emph {et~al.}(2021)\citenamefont {Wu},
  \citenamefont {Ray}, \citenamefont {Zhao}, \citenamefont {Sun},\ and\
  \citenamefont {Rebentrost}}]{wu2021quantum}%
  \BibitemOpen
  \bibfield  {author} {\bibinfo {author} {\bibfnamefont {B.}~\bibnamefont
  {Wu}}, \bibinfo {author} {\bibfnamefont {M.}~\bibnamefont {Ray}}, \bibinfo
  {author} {\bibfnamefont {L.}~\bibnamefont {Zhao}}, \bibinfo {author}
  {\bibfnamefont {X.}~\bibnamefont {Sun}}, \ and\ \bibinfo {author}
  {\bibfnamefont {P.}~\bibnamefont {Rebentrost}},\ }\href@noop {} {\bibfield
  {journal} {\bibinfo  {journal} {Physical Review A}\ }\textbf {\bibinfo
  {volume} {103}},\ \bibinfo {pages} {042422} (\bibinfo {year}
  {2021})}\BibitemShut {NoStop}%
\bibitem [{\citenamefont {Xu}\ \emph {et~al.}(2022)\citenamefont {Xu},
  \citenamefont {Zhang}, \citenamefont {Liang}, \citenamefont {Wang},
  \citenamefont {Li}, \citenamefont {Jian},\ and\ \citenamefont
  {Shen}}]{xu2022variational}%
  \BibitemOpen
  \bibfield  {author} {\bibinfo {author} {\bibfnamefont {L.}~\bibnamefont
  {Xu}}, \bibinfo {author} {\bibfnamefont {X.-Y.}\ \bibnamefont {Zhang}},
  \bibinfo {author} {\bibfnamefont {J.-M.}\ \bibnamefont {Liang}}, \bibinfo
  {author} {\bibfnamefont {J.}~\bibnamefont {Wang}}, \bibinfo {author}
  {\bibfnamefont {M.}~\bibnamefont {Li}}, \bibinfo {author} {\bibfnamefont
  {L.}~\bibnamefont {Jian}}, \ and\ \bibinfo {author} {\bibfnamefont {S.-q.}\
  \bibnamefont {Shen}},\ }\href@noop {} {\bibfield  {journal} {\bibinfo
  {journal} {Communications in Theoretical Physics}\ }\textbf {\bibinfo
  {volume} {74}},\ \bibinfo {pages} {055106} (\bibinfo {year}
  {2022})}\BibitemShut {NoStop}%
\bibitem [{\citenamefont {Patti}\ \emph {et~al.}(2023)\citenamefont {Patti},
  \citenamefont {Kossaifi}, \citenamefont {Anandkumar},\ and\ \citenamefont
  {Yelin}}]{patti2023quantum}%
  \BibitemOpen
  \bibfield  {author} {\bibinfo {author} {\bibfnamefont {T.~L.}\ \bibnamefont
  {Patti}}, \bibinfo {author} {\bibfnamefont {J.}~\bibnamefont {Kossaifi}},
  \bibinfo {author} {\bibfnamefont {A.}~\bibnamefont {Anandkumar}}, \ and\
  \bibinfo {author} {\bibfnamefont {S.~F.}\ \bibnamefont {Yelin}},\ }\href@noop
  {} {\bibfield  {journal} {\bibinfo  {journal} {Quantum}\ }\textbf {\bibinfo
  {volume} {7}},\ \bibinfo {pages} {1057} (\bibinfo {year} {2023})}\BibitemShut
  {NoStop}%
\bibitem [{\citenamefont {Aharonov}\ \emph {et~al.}(2006)\citenamefont
  {Aharonov}, \citenamefont {Jones},\ and\ \citenamefont
  {Landau}}]{aharonov2006polynomial}%
  \BibitemOpen
  \bibfield  {author} {\bibinfo {author} {\bibfnamefont {D.}~\bibnamefont
  {Aharonov}}, \bibinfo {author} {\bibfnamefont {V.}~\bibnamefont {Jones}}, \
  and\ \bibinfo {author} {\bibfnamefont {Z.}~\bibnamefont {Landau}},\ }in\
  \href@noop {} {\emph {\bibinfo {booktitle} {Proceedings of the thirty-eighth
  annual ACM symposium on Theory of computing}}}\ (\bibinfo {year} {2006})\
  pp.\ \bibinfo {pages} {427--436}\BibitemShut {NoStop}%
\bibitem [{\citenamefont {Greenbaum}(2015)}]{greenbaum2015introduction}%
  \BibitemOpen
  \bibfield  {author} {\bibinfo {author} {\bibfnamefont {D.}~\bibnamefont
  {Greenbaum}},\ }\href@noop {} {\bibfield  {journal} {\bibinfo  {journal}
  {arXiv preprint arXiv:1509.02921}\ } (\bibinfo {year} {2015})}\BibitemShut
  {NoStop}%
\bibitem [{\citenamefont {Torlai}\ \emph {et~al.}(2018)\citenamefont {Torlai},
  \citenamefont {Mazzola}, \citenamefont {Carrasquilla}, \citenamefont
  {Troyer}, \citenamefont {Melko},\ and\ \citenamefont
  {Carleo}}]{torlai2018neural}%
  \BibitemOpen
  \bibfield  {author} {\bibinfo {author} {\bibfnamefont {G.}~\bibnamefont
  {Torlai}}, \bibinfo {author} {\bibfnamefont {G.}~\bibnamefont {Mazzola}},
  \bibinfo {author} {\bibfnamefont {J.}~\bibnamefont {Carrasquilla}}, \bibinfo
  {author} {\bibfnamefont {M.}~\bibnamefont {Troyer}}, \bibinfo {author}
  {\bibfnamefont {R.}~\bibnamefont {Melko}}, \ and\ \bibinfo {author}
  {\bibfnamefont {G.}~\bibnamefont {Carleo}},\ }\href@noop {} {\bibfield
  {journal} {\bibinfo  {journal} {Nature Physics}\ }\textbf {\bibinfo {volume}
  {14}},\ \bibinfo {pages} {447} (\bibinfo {year} {2018})}\BibitemShut
  {NoStop}%
\bibitem [{\citenamefont {Cramer}\ \emph {et~al.}(2010)\citenamefont {Cramer},
  \citenamefont {Plenio}, \citenamefont {Flammia}, \citenamefont {Somma},
  \citenamefont {Gross}, \citenamefont {Bartlett}, \citenamefont
  {Landon-Cardinal}, \citenamefont {Poulin},\ and\ \citenamefont
  {Liu}}]{cramer2010efficient}%
  \BibitemOpen
  \bibfield  {author} {\bibinfo {author} {\bibfnamefont {M.}~\bibnamefont
  {Cramer}}, \bibinfo {author} {\bibfnamefont {M.~B.}\ \bibnamefont {Plenio}},
  \bibinfo {author} {\bibfnamefont {S.~T.}\ \bibnamefont {Flammia}}, \bibinfo
  {author} {\bibfnamefont {R.}~\bibnamefont {Somma}}, \bibinfo {author}
  {\bibfnamefont {D.}~\bibnamefont {Gross}}, \bibinfo {author} {\bibfnamefont
  {S.~D.}\ \bibnamefont {Bartlett}}, \bibinfo {author} {\bibfnamefont
  {O.}~\bibnamefont {Landon-Cardinal}}, \bibinfo {author} {\bibfnamefont
  {D.}~\bibnamefont {Poulin}}, \ and\ \bibinfo {author} {\bibfnamefont {Y.-K.}\
  \bibnamefont {Liu}},\ }\href@noop {} {\bibfield  {journal} {\bibinfo
  {journal} {Nature communications}\ }\textbf {\bibinfo {volume} {1}},\
  \bibinfo {pages} {149} (\bibinfo {year} {2010})}\BibitemShut {NoStop}%
\bibitem [{\citenamefont {Mohseni}\ \emph {et~al.}(2008)\citenamefont
  {Mohseni}, \citenamefont {Rezakhani},\ and\ \citenamefont
  {Lidar}}]{mohseni2008quantum}%
  \BibitemOpen
  \bibfield  {author} {\bibinfo {author} {\bibfnamefont {M.}~\bibnamefont
  {Mohseni}}, \bibinfo {author} {\bibfnamefont {A.~T.}\ \bibnamefont
  {Rezakhani}}, \ and\ \bibinfo {author} {\bibfnamefont {D.~A.}\ \bibnamefont
  {Lidar}},\ }\href@noop {} {\bibfield  {journal} {\bibinfo  {journal}
  {Physical Review A}\ }\textbf {\bibinfo {volume} {77}},\ \bibinfo {pages}
  {032322} (\bibinfo {year} {2008})}\BibitemShut {NoStop}%
\bibitem [{\citenamefont {Altepeter}\ \emph {et~al.}(2003)\citenamefont
  {Altepeter}, \citenamefont {Branning}, \citenamefont {Jeffrey}, \citenamefont
  {Wei}, \citenamefont {Kwiat}, \citenamefont {Thew}, \citenamefont
  {O’Brien}, \citenamefont {Nielsen},\ and\ \citenamefont
  {White}}]{altepeter2003ancilla}%
  \BibitemOpen
  \bibfield  {author} {\bibinfo {author} {\bibfnamefont {J.~B.}\ \bibnamefont
  {Altepeter}}, \bibinfo {author} {\bibfnamefont {D.}~\bibnamefont {Branning}},
  \bibinfo {author} {\bibfnamefont {E.}~\bibnamefont {Jeffrey}}, \bibinfo
  {author} {\bibfnamefont {T.}~\bibnamefont {Wei}}, \bibinfo {author}
  {\bibfnamefont {P.~G.}\ \bibnamefont {Kwiat}}, \bibinfo {author}
  {\bibfnamefont {R.~T.}\ \bibnamefont {Thew}}, \bibinfo {author}
  {\bibfnamefont {J.~L.}\ \bibnamefont {O’Brien}}, \bibinfo {author}
  {\bibfnamefont {M.~A.}\ \bibnamefont {Nielsen}}, \ and\ \bibinfo {author}
  {\bibfnamefont {A.~G.}\ \bibnamefont {White}},\ }\href@noop {} {\bibfield
  {journal} {\bibinfo  {journal} {Physical Review Letters}\ }\textbf {\bibinfo
  {volume} {90}},\ \bibinfo {pages} {193601} (\bibinfo {year}
  {2003})}\BibitemShut {NoStop}%
\bibitem [{\citenamefont {Lanyon}\ \emph {et~al.}(2017)\citenamefont {Lanyon},
  \citenamefont {Maier}, \citenamefont {Holz{\"a}pfel}, \citenamefont
  {Baumgratz}, \citenamefont {Hempel}, \citenamefont {Jurcevic}, \citenamefont
  {Dhand}, \citenamefont {Buyskikh}, \citenamefont {Daley}, \citenamefont
  {Cramer} \emph {et~al.}}]{lanyon2017efficient}%
  \BibitemOpen
  \bibfield  {author} {\bibinfo {author} {\bibfnamefont {B.}~\bibnamefont
  {Lanyon}}, \bibinfo {author} {\bibfnamefont {C.}~\bibnamefont {Maier}},
  \bibinfo {author} {\bibfnamefont {M.}~\bibnamefont {Holz{\"a}pfel}}, \bibinfo
  {author} {\bibfnamefont {T.}~\bibnamefont {Baumgratz}}, \bibinfo {author}
  {\bibfnamefont {C.}~\bibnamefont {Hempel}}, \bibinfo {author} {\bibfnamefont
  {P.}~\bibnamefont {Jurcevic}}, \bibinfo {author} {\bibfnamefont
  {I.}~\bibnamefont {Dhand}}, \bibinfo {author} {\bibfnamefont
  {A.}~\bibnamefont {Buyskikh}}, \bibinfo {author} {\bibfnamefont
  {A.}~\bibnamefont {Daley}}, \bibinfo {author} {\bibfnamefont
  {M.}~\bibnamefont {Cramer}},  \emph {et~al.},\ }\href@noop {} {\bibfield
  {journal} {\bibinfo  {journal} {Nature Physics}\ }\textbf {\bibinfo {volume}
  {13}},\ \bibinfo {pages} {1158} (\bibinfo {year} {2017})}\BibitemShut
  {NoStop}%
\bibitem [{\citenamefont {Blume-Kohout}\ \emph {et~al.}(2013)\citenamefont
  {Blume-Kohout}, \citenamefont {Gamble}, \citenamefont {Nielsen},
  \citenamefont {Mizrahi}, \citenamefont {Sterk},\ and\ \citenamefont
  {Maunz}}]{blume2013robust}%
  \BibitemOpen
  \bibfield  {author} {\bibinfo {author} {\bibfnamefont {R.}~\bibnamefont
  {Blume-Kohout}}, \bibinfo {author} {\bibfnamefont {J.~K.}\ \bibnamefont
  {Gamble}}, \bibinfo {author} {\bibfnamefont {E.}~\bibnamefont {Nielsen}},
  \bibinfo {author} {\bibfnamefont {J.}~\bibnamefont {Mizrahi}}, \bibinfo
  {author} {\bibfnamefont {J.~D.}\ \bibnamefont {Sterk}}, \ and\ \bibinfo
  {author} {\bibfnamefont {P.}~\bibnamefont {Maunz}},\ }\href@noop {}
  {\bibfield  {journal} {\bibinfo  {journal} {arXiv preprint arXiv:1310.4492}\
  } (\bibinfo {year} {2013})}\BibitemShut {NoStop}%
\bibitem [{\citenamefont {Cowling}\ and\ \citenamefont
  {Price}(1984)}]{cowling1984bandwidth}%
  \BibitemOpen
  \bibfield  {author} {\bibinfo {author} {\bibfnamefont {M.~G.}\ \bibnamefont
  {Cowling}}\ and\ \bibinfo {author} {\bibfnamefont {J.~F.}\ \bibnamefont
  {Price}},\ }\href@noop {} {\bibfield  {journal} {\bibinfo  {journal} {SIAM
  journal on mathematical analysis}\ }\textbf {\bibinfo {volume} {15}},\
  \bibinfo {pages} {151} (\bibinfo {year} {1984})}\BibitemShut {NoStop}%
\end{thebibliography}%

\onecolumngrid

\begin{appendices}

\appendix
\section{Completeness Proof}
\label{Appendix A: Completeness Proof}

In the main text, the calculation of $\Tr\{\rho^k\}$ has been mathematically transformed into investigating $\Tr\{G^k\}$, where $\Tr\{G^k\}$ represents the trace of $G^k$, which is obtained by decomposing the random state into pure states and then taking a weighted average. Therefore, the primary focus is on calculating $\Tr\{\mathcal{G}_q^k\}$.

After selecting a combination ${\mathcal{G}_q^k: G_{q_1}G_{q_2}\cdots G_{q_t}\cdots G_{q_k}}$, it is necessary to determine the dimension of the subspace $d$. For the subspace determined by the combination $\mathcal{G}_q^k$, if the dimension of the subspace is $d$, then the quantum states $\rho{s}$ prepared by different random gates of dimension $d$, where $s \in \{1, 2, \ldots, d\}$, are guaranteed to be linearly independent.Preparation of $d$ linearly independent quantum states using $d$ $n$-bit random gates: $\rho_s = U_s|0\rangle\langle0|^{\otimes n}U_s^\dagger$. The random gate $U_s$ belongs to different random gates within a certain combination $\mathcal{G}_q^k$.

Preparation of the remaining $d^2-d$ quantum states: Assuming we have already prepared $d$ linearly independent quantum states: $\rho_1, \rho_2,\ldots,\rho_d$, we need to prepare an additional $d^2-d$ quantum states to form a complete $d^2$-dimensional inner product space. In a finite-dimensional linear space, it is sufficient for all $d^2$ quantum states to be linearly independent from each other.

For the quantum state $\rho_s=U_s|0\rangle\langle 0|^{\otimes n}U_s^\dagger$, where $s,s^\prime \in\{1,2,\ldots,d\},s\neq s^\prime$, we apply the quantum gate $G_{s^\prime}=U_{s^\prime}{G_0}U_{s^\prime}^\dagger=I-2\rho_{s^\prime}$ to obtain:
\begin{equation}
\begin{aligned}
\rho_{ss^\prime}&={G}_{s^\prime}\rho_s{G}_{s^\prime}^\dagger=(I-2\rho_{s^\prime})\rho_{s}(I-2\rho_{s^\prime})\\
&=\rho_s-2\rho_{s^\prime}\rho_s-2\rho_s\rho_{s^\prime}+4\rho_{s^\prime}\rho_s\rho_{s^\prime}.\\
\end{aligned}
\end{equation}
For the corresponding $\rho_{s^\prime s}$, we have:
\begin{equation}
\rho_{s^\prime s}=\rho_{s^\prime}-2\rho_{s^\prime}\rho_s-2\rho_s\rho_{s^\prime}+4\rho_s\rho_{s^\prime}\rho_{s}.
\end{equation}
It can be shown that the quantum states $\rho_s,\rho_{s^\prime},\rho_{ss^\prime},\rho_{s^\prime s}$ are linearly dependent. If we construct all quantum states using this method, then these quantum states will be linearly dependent and cannot form a complete $d^2$-dimensional inner product space.

To ensure linear independence, we modify the gate $G_0$ to $G_\theta$, where $\theta\neq k\pi, k\in\mathbb{Z}$:
\begin{equation}    
G_0 =\begin{pmatrix} -1 & 0 & \cdots & 0 \\
    0 & 1& \cdots & 0 \\
    \vdots & \vdots & \ddots & \vdots \\
    0 & 0 & \cdots & 1\end{pmatrix} \longrightarrow G_\theta =\begin{pmatrix} e^{i\theta} & 0 & \cdots & 0 \\
    0 & 1& \cdots & 0 \\
    \vdots & \vdots & \ddots & \vdots \\
    0 & 0 & \cdots & 1\end{pmatrix}  =\begin{pmatrix} 1 & 0 & \cdots & 0 \\
    0 & 1& \cdots & 0 \\
    \vdots & \vdots & \ddots & \vdots \\
    0 & 0 & \cdots & 1\end{pmatrix} -\begin{pmatrix} 1-e^{i\theta}& 0 & \cdots & 0 \\
    0 & 1& \cdots & 0 \\
    \vdots & \vdots & \ddots & \vdots \\
    0 & 0 & \cdots & 1\end{pmatrix}
\end{equation}
After this change, the transformed${G}_{s}(\theta) = {U}_{s} G_\theta {U}_{s}^\dagger$. We represent $G_s({\theta})={U}_{s}G({\theta}){U}_{s}^\dagger$ as ${G}_{s}(\theta)=I-\rho_{s}(\theta)$, and $\rho_s(\theta)\neq\rho_{s}$. Following the same approach as before, we prepare the quantum states:
\begin{equation}
\begin{aligned}
&1.\rho_s={U}_s|0\rangle\langle 0|^{\otimes n} {U}_s^\dagger,\\
&2.\rho_{s^\prime}={U}_{s^\prime}|0\rangle\langle 0|^{\otimes n} {U}_{s^\prime}^\dagger,\\
&3.\rho_{ss^\prime}=(I-2\rho_{s^\prime})\rho_{s}(I-2\rho_{s^\prime})\\ &\ \ \ \ \ \ \ \ \ =\rho_s-2\rho_{s^\prime}\rho_s-2\rho_s\rho_{s^\prime}+4\rho_{s^\prime}\rho_s\rho_{s^\prime},\\
&4.\rho_{s^\prime s}(\theta)={G}_{s}(\theta)\rho_{s^\prime}{G}_{s}(\theta)^\dagger\\
&\ \ \ \ \ \ \ \ \ \ \ \ \ \  =(I-\rho_{s}(\theta))\rho_{s^\prime}(I-\rho_{s}(\theta))\\
&\ \ \ \ \ \ \ \ \ \ \ \ \ \  =\rho_{s^\prime}-\rho_{s^\prime}\rho_s(\theta)-\rho_{s}(\theta)\rho_{s^\prime}+\rho_{s}(\theta)\rho_{s^\prime}\rho_{s}(\theta).
\end{aligned}
\end{equation}
From the above equation, we observe that rotating the quantum gate $G_0$ by an arbitrary angle $\theta\neq k\pi, k=0,1,2\cdots$ changes the originally linearly dependent quantum states, prepared using $G_0$, into linearly independent ones.

By using similar methods, it is possible to prepare the remaining $d(d-1)$ quantum states and ensure that they are linearly independent. This allows us to construct a complete $d^2$-dimensional linear space for performing a full tomography of the PTM
 under study.

\section{More Numerical Simulation}
\label{Appendix B: More Numerical Simulation}
\addcontentsline{toc}{section}{Appendix B: More Numerical Simulation}

For the Hadamard Test procedure, we incorporate noise simulation into its measurement process by injecting random numbers drawn from a Gaussian distribution with a standard deviation of 0.01. The graphical representation below illustrates the outcomes prior to and subsequent to the integration of these random numbers:
\begin{figure}[H]
        \centering
        \includegraphics[width=.8\linewidth]{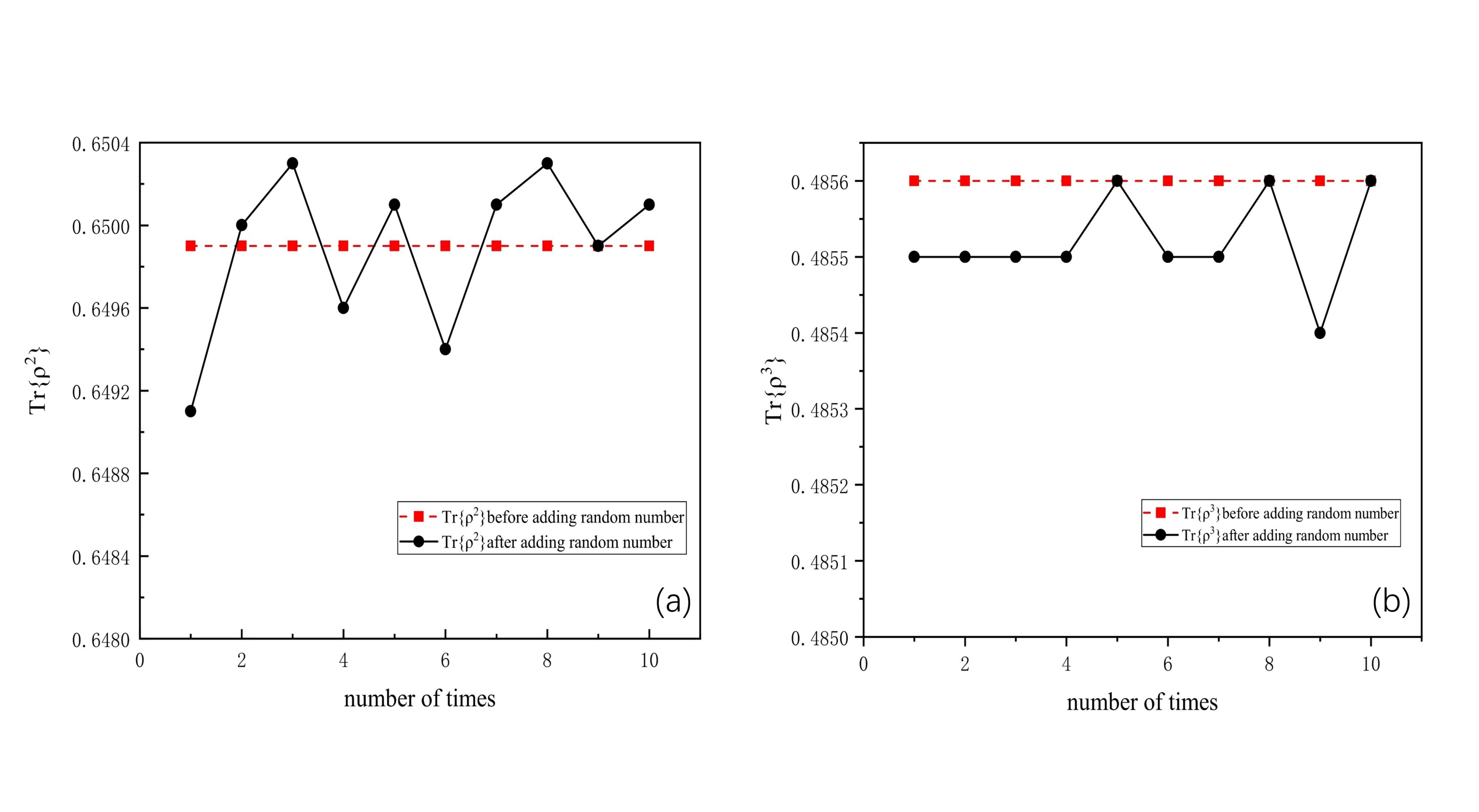}
        \caption{This figure presents numerical results for $\Tr\{\rho^2\}$ and $\Tr\{\rho^3\}$, taking into account noise, using the Hadamard Test method. Subfigure (a) displays the numerical results for $\Tr\{\rho^2\}$, and subfigure (b) shows the numerical results for $\Tr\{\rho^3\}$. The black lines represent the results after adding random numbers, while the red lines represent the results before adding random gates.}
        \label{fig:your_label1}
\end{figure}
For Tomography, we introduce noise simulation by adding random numbers to its PTM elements as well as g-matrix elements. Since the singular values of the $g$-matrix are small, we choose random numbers sampled from a Gaussian distribution with a standard deviation of 0.0001 to simulate the noise. The results before and after adding the random numbers are shown in the graph below:
\begin{figure}[H]
        \centering
        \includegraphics[width=.8\linewidth]{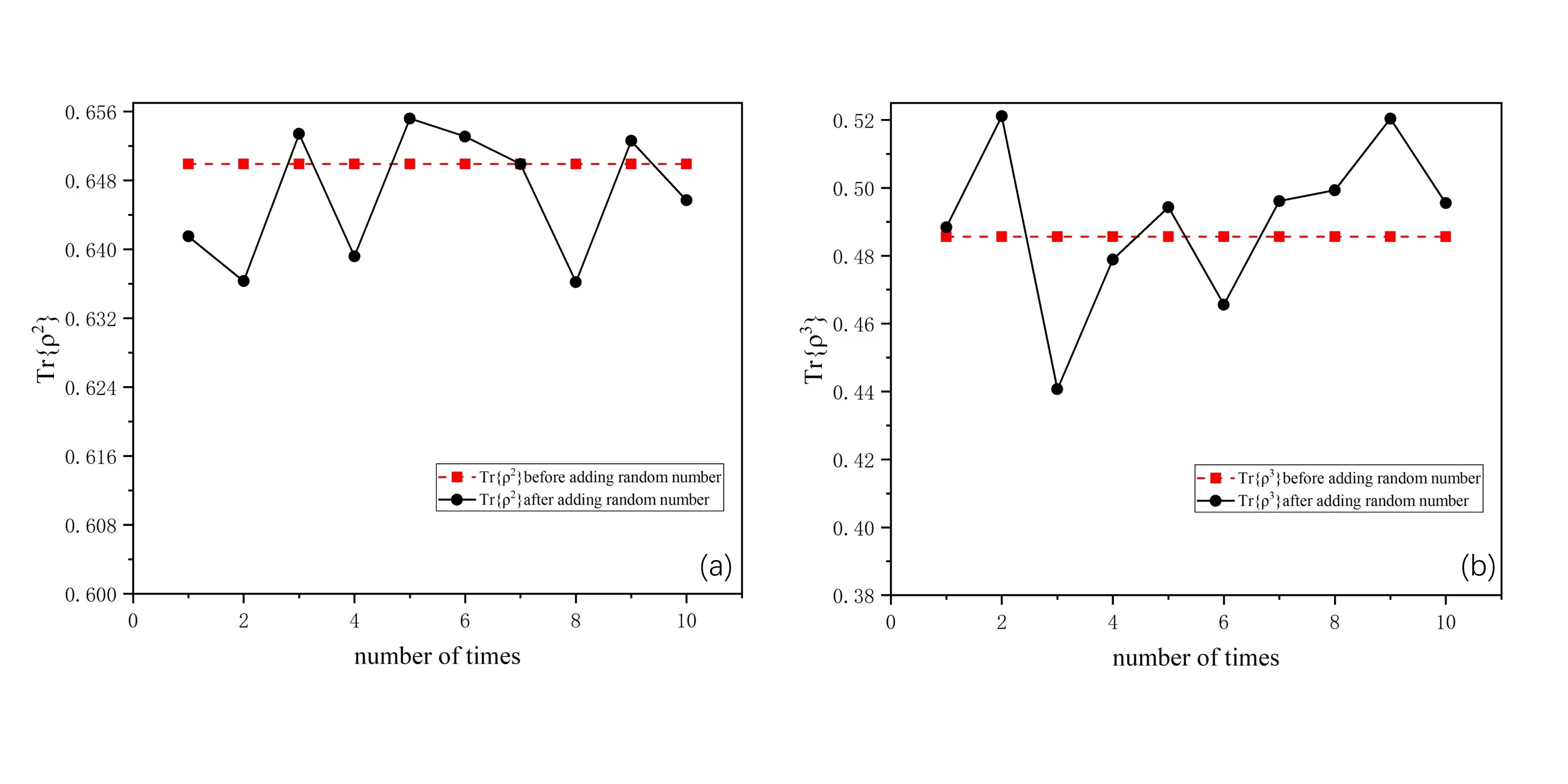}
        \caption{This figure showcases the numerical outcomes of $\Tr\{\rho^2\}$ and $\Tr\{\rho^3\}$ calculated using the GST  method while accounting for noise. Subfigure (a) presents the numerical results for $\Tr\{\rho^2\}$, and subfigure (b) illustrates the numerical results for $\Tr\{\rho^3\}$. The black lines denote the results obtained after the addition of random numbers, while the red lines represent the results before the introduction of random gates.}
        \label{fig:your_label2}
\end{figure}

Derived from the data depicted in the provided graph, we are able to distinctly discern the repercussions stemming from the introduction or absence of noise. When noise is absent, the output data from the quantum circuit manifests as consistently stable and precise. Nonetheless, the scenario takes a discernible turn upon the infusion of noise. The introduction of noise precipitates volatility in the quantum circuit's outcomes, potentially culminating in errors. Noise can engender an unreliable exchange of information amidst quantum bits, thereby injecting substantial ambiguity into the computational process. These sources of noise might encompass phenomena such as dephasing, errors in quantum gate operations, imprecisions in measurements, or other external disturbances.

In real-world scenarios, noise presents a significant hurdle for the advancement of quantum computing and quantum information processing. To counteract the disruptive influence of noise, researchers are tirelessly engaged in refining quantum error correction and noise suppression techniques. The primary goal of these approaches is to bolster the resilience of quantum circuits and heighten the precision of outcomes, ultimately striving for a more dependable realm of quantum computing.Furthermore, the ramifications of noise can exhibit variability across diverse quantum algorithms and tasks. As a result, practical applications necessitate a thorough assessment of algorithm robustness alongside the prevailing noise levels. Achieving equilibrium between these considerations is crucial in determining the most optimal course of action.

\section{Subspace dimension}
In this section, we study the dimensions of the invariant subspaces corresponding to nonlinear functions of different types of density matrices.
Let's start with the simplest case, $\Tr\{\rho^m\}$. In our algorithm, we decompose the computation of $\Tr\{\rho^m\}$ into the computation of $\Tr\{G_{q_1} G_{q_2}\dots G_{q_k}\}$, for $k = 0 ,1, \dots, m $. It is obvious that $\textbf{span}(|\psi_{q_1}\rangle, |\psi_{q_2}\rangle,\dots,|\psi_{q_k}\rangle)$ forms a non-trivial invariant subspace of $G_{q_1} G_{q_2}\dots G_{q_k}$. Therefore, the largest non-trivial invariant subspace involved in the calculation of $\Tr\{\rho^m\}$ is m-dimensional.

A more general nonlinear function is $\Tr\{P_1\rho_1 P_2\rho_2\dots P_m\rho_m\}$. We can rewrite this function as $\Tr\{P^{\prime}\rho_1^{\prime} \rho_2^{\prime}\dots \rho_m^{\prime}\}$, where $\rho_i^{\prime} = (\Pi_{l=1}^i P_l)\rho_n(\Pi_{l=1}^i P_l)^{\dagger} $ and $P^{\prime} =\Pi_{l=1}^i P_l $. Remark that although $\rho_1^{\prime} \neq \rho_2^{\prime} \neq \dots \neq \rho_m^{\prime}$, the algorithms we proposed in this paper can also be used.
 During the calculation, we need to calculate $\Tr\{P^{\prime} G_{q_1} G_{q_2}\dots G_{q_k}\}$, for $k = 0 ,1 \dots, m $, where $P^{\prime}$ is a Pauli operator. Note that $G_{q_i}$ acts on $|\psi\rangle$ to get a superposition of $|\psi_{q_i}\rangle$ and $|\psi\rangle$. Thus the non-trivial invariant subspace corresponding to $P  G_{q_1} G_{q_2}\dots G_{q_k}$ is $\textbf{span}(|\psi_{q_1}\rangle,\dots,|\psi_{q_m}\rangle,P|\psi_{q_1}\rangle,\dots,P|\psi_{q_m}\rangle)$, of which the largest possible dimension is $2m$ dimensions.

\end{appendices}

\end{document}